\documentclass[aps,pre,reprint]{revtex4-2}
\usepackage[latin1]{inputenc}
\usepackage{amsmath}
\usepackage{amsfonts}
\usepackage{amssymb}
\usepackage[dvipdfmx]{graphicx}
\usepackage{xcolor}
\usepackage{bm}
\usepackage[a4paper,top=2cm,bottom=2cm,left=2cm,right=2cm]{geometry}

\begin{document}
	
\author{Takahiro Sakaue}
\affiliation{Department of Physical Sciences, Aoyama Gakuin University, 5-10-1 Fuchinobe, Chuo-ku, Sagamihara, Japan}
\email{sakaue@phys.aoyamna.ac.jp}
\author{Enrico Carlon}
\affiliation{Soft Matter and Biophysics, KU Leuven, Celestijnenlaan 200D, B-3001, Leuven, Belgium}

\title{Physics of active polymers: scaling analysis via a compounding formula}

\begin{abstract}
Active polymeric systems exhibit a rich spectrum of non-equilibrium phenomena 
arising from stochastic forces that explicitly break detailed balance. Despite the 
rapid growth of experimental and numerical studies, analytical progress remains 
limited. To date, theoretical understanding relies largely on variants of the active 
Rouse model, whose formal solutions, though exact, are often obscured by summations 
over Rouse modes and therefore provide limited direct physical insight.
In this work, we develop a transparent scaling theory that captures the tagged-monomer 
mean-squared displacement (MSD) in active polymers through a compounding formula: the MSD 
of a monomer in the chain is expressed as that of an isolated active particle, modulated 
by a connectivity factor encoding tension propagation along the polymer backbone. This approach 
isolates the role of activity from that of polymer connectivity and reveals the emergent 
dynamical regimes in a physically intuitive manner. We test the scaling predictions against 
exact calculations for a broad class of generalized active polymer models driven by diverse 
noise statistics. The agreement demonstrates the robustness of the scaling framework across 
microscopic details. Our results provide a simple and extensible theoretical structure 
that can be applied to complex and analytically intractable active polymer systems, thereby 
offering a unifying perspective on non-equilibrium polymer dynamics.
\end{abstract}

\maketitle

\section{Introduction}

Active polymeric systems are driven out of equilibrium by forces of non-thermal origin, 
arising from internal energy-consuming processes. In living matter, there are numerous 
examples of such systems, including cytoskeletal filaments such as actin and 
microtubules \cite{mizno07,marc13,dema15,juli18,eise17,Gompper_2020,bane20}. 
These polymers are persistently maintained far from equilibrium by ATP-dependent mechanisms 
that convert chemical energy into mechanical work, giving rise to a wide variety of 
nonequilibrium behaviors \cite{Lowen_2014,Shin_2015,vand15,Sakaue_2017,vand17,osma17,Put_2019,bian18,smre20,sala22,li23,Goychuk23,malg25}.
A paradigmatic example of an active polymer is chromatin, the DNA-protein complex that 
occupies the nucleus of eukaryotic cells. Chromatin is subject to a variety of active 
processes, including transcription, replication, chromatin remodeling, and loop extrusion, 
all of which reshape its organization in space and time~\cite{forte26}. As a result, chromatin provides 
a key natural platform for studying how active forces influence polymer structure and 
dynamics \cite{webe12,tort20,Dudko_2019,garini24,yuan24}. 
The latter is commonly studied by the time evolution of the mean-squared 
displacement (MSD) of a tagged genomic locus. In many experiments, the MSD 
displays an anomalous power-law scaling of the form
\begin{eqnarray}
\langle \Delta \vec{r}^{\ 2}(t) \rangle \sim t^{\alpha},    
\end{eqnarray}
where the exponent deviates from the value $\alpha = 1$ expected for simple 
Brownian motion. 
A wide range of values for the anomalous diffusion exponent $\alpha$ has 
been reported across different cell types \cite{webe10,webe12,hajj13,back15,wang15,Kimura_2022,Maeshima_2020}.  
Early measurements of chromosomal locus dynamics in bacteria revealed subdiffusive 
behavior characterized by an exponent $\alpha \approx 0.4$ \cite{webe10}.  
In other organisms, such as yeast, experimental data yielded values consistent 
with $\alpha = 1/2$, in agreement with the predictions of the equilibrium Rouse 
model for polymer dynamics \cite{soco19}.  
In a series of E. coli chromosomal tracking experiments \cite{jave14}, 
most loci displayed subdiffusive motion, while a distinct subset exhibited 
superdiffusive trajectories approaching ballistic behavior, corresponding 
to an effective exponent $\alpha \approx 2$.

From the theoretical side, an exactly solvable model of chromatin dynamics 
is the active Rouse model, which exhibits a rich spectrum of dynamical behaviors, 
including both subdiffusive and superdiffusive regimes \cite{vand15,vand17,Gompper_2020,eise17,Put_2019,osma17}. 
However, despite the 
analytical tractability and the many insights these exact solutions provide, the resulting 
expressions for observables such as the mean-squared displacement (MSD) typically 
involve sums over Rouse modes. As a consequence, one must usually resort to numerical 
evaluation of these mode sums, which obscures the physical origin of dynamical 
crossovers and the mechanisms underlying changes of scaling regimes.

The aim of this paper is to develop a scaling theory of active polymer dynamics that 
overcomes these limitations. Our approach provides simple, physically transparent predictions 
for the MSD and related observables, clarifying how different types of active forcing and 
polymer properties control the emergence of subdiffusive, diffusive, and superdiffusive 
behaviors \cite{saka26_lett}. In doing so, it not only rationalizes the results obtained 
from exactly solvable 
active Rouse models, but also enables systematic exploration of more complex situations 
that are not accessible to exact analytical treatments. In particular, we consider
polymers with additional interactions beyond the ideal Rouse description, which may 
serve as a model of crumpled globule with an appropriate choice on the form of long 
range interactions.
The central concept in our work is a compounding formula, which expresses the 
tagged monomer MSD as that of unconnected/isolated monomer and the domain size 
$m(\tau)$ for the cooperative motion over the time scale $\tau$, thereby decomposing
the monomer dynamics into two separate contributions which are easier to infer from
scaling arguments.

This paper is organized as follows. In Sec.~\ref{sec:compounding}, we introduce 
the compounding formula and clarify its meaning with insights from polymer physics 
perspectives. The isolated monomer MSD and the domain size $m(\tau)$ are combined 
in Sec.~\ref{scaling_prediction} to produce the scaling prediction for the 
tagged monomer MSD. Here we will also introduce the two protocols to measure 
the MSD, i.e., the steady state and the transient. 
We then present the rigorous calculation in Sec.~\ref{sec:rigorous_approach} by 
employing (i) the normal mode analysis in Sec.~\ref{sec:normal_mode_analysis} and 
(ii) the real space analysis in Sec.~\ref{sec:real_space_analysis}.
The rigorous results allow us to verify the validity of the compounding formula
approach for a broad range of cases and active noises, both exponentially and power-law
correlated. As a straightforward extension of the real space analysis, we report in 
Sec.~\ref{sec:H} the calculations of the displacement correlation of two monomers. 
This quantity underpins a distinction between steady-state and transient, and 
clarifies the hitherto overlooked mechanism for the steady-state MSD superdiffusive 
scaling. Finally, Section~\ref{sec:conclusion} summarizes the main results and 
discusses possible extensions for future work. 
A concise presentation of the key findings of this work, emphasizing their physical 
implications while omitting the technical derivations, is provided in Ref.~\cite{saka26_lett}.

\section{The compounding formula}
\label{sec:compounding}
In the Rouse description, the dynamics of $z(n,t)$, a cartesian component of
the position of the monomer $n$ at time $t$ is described by the following Langevin 
equation
\begin{eqnarray}
\gamma \frac{\partial z(n,t)}{\partial t} = 
k \frac{\partial^2 z(n,t)}{\partial n^2} + f(n,t)
\label{eq:Rouse}
\end{eqnarray}
where $\gamma$ is the friction factor, $k$ is the spring constant and $f(n,t)$ is the 
zero-mean random force or noise acting on $n$-th monomer. Here we employ a continuum 
description, hence, $n \in[0,N]$ is a continuous variable, which would be valid for 
length scale larger than the monomer size. The models described by Eq.~\eqref{eq:Rouse} 
are distinguished according to the statistics of the noise. Writing the autocorrelation 
of the noise
\begin{eqnarray}
\langle f(n,t) f(n',t') \rangle = A g(|t-t'|) \delta(n-n'), 
\label{auto_corr_noise}
\end{eqnarray}
we have the following classification.
\begin{itemize}
\item[(a)] 
If the relations
\begin{eqnarray}
A = 2 \gamma k_BT, g(t) = \delta (t)
\label{FDT}
\end{eqnarray}
are fulfilled, Eq.~\eqref{auto_corr_noise} amounts to the fluctuation-dissipation 
theorem, hence, $f(n,t)=f_{th}(n,t)$ is regarded as the thermal noise, where $k_BT$ 
is the thermal energy. We call Eq.~\eqref{eq:Rouse} with the thermal noise 
``equilibrium Rouse model".
\item[(b)] 
If either one of relations~\eqref{FDT} is violated, $f(n,t)$ is no longer regarded 
as thermal origin. Since the resulting non-equilibrium dynamics are diverse depending 
on how the relations~\eqref{FDT} is violated, it is convenient to introduce 
the following subclasses.

\begin{itemize}
\item[(b-1)] 
If $A \neq 2 \gamma k_BT$, but the noise is still white, i.e., $g(t) = \delta (t)$, many 
observables such as MSD can be described by the equilibrium Rouse model by introducing 
the effective temperature $T_{eff} = A/(2 \gamma k_B)$. Such a model may be thus called 
still an equilibrium model in an effective sense. However,  the fact $T_{eff} \neq T$ 
implies the presence of heat flow in the system, and the extra thermodynamic cost to 
maintain such a non-equilibrium state.

\item[(b-2)]
If $g(t) \neq \delta (t)$, i.e., the noise is colored and has some persistence, it 
gives rise to genuine non-equilibrium behaviors. For a simple particle such as colloid, 
the models kicked by such persistent noises, often called active particles, have been 
extensively studied. Here, we call their polymeric extension active polymers, which 
exhibit nontrivial dynamics apparently different from those in the equilibrium Rouse model.
\end{itemize}

Since the thermal noise $f_{th}(n,t)$ is ubiquitous in small systems, the non-thermal 
noise may be present on top of it. In that case, assuming the independence of these 
two noises, $f(n,t)$ in the (b-1) case is regarded as the sum of these noises, thus 
one usually expects $T_{eff} > T$. In the case (b-2), the behavior of the model is 
a superposition of the thermal behavior due to $f_{th}(n,t)$ and the active behavior 
due to $f(n,t)$. Which contribution dominates in the observed behavior depends 
on the magnitude of the active noise $A$ and the time-scale, hence, affected by the 
persistence encoded in $g(t)$. 
\end{itemize}

Here, we develop a scaling theory to capture the dynamics of active polymers, i.e., the case (b-2).
Our scaling analysis relies on expressing $\langle \Delta z^2(n,\tau) \rangle$ the MSD of 
the monomer $n$ of the chain at time-scale $\tau$ as the ratio of the MSD of an isolated monomer 
$\langle \Delta z_i^2(\tau) \rangle$ with the number of dynamically connected monomers 
$m(\tau)$ as follows
\begin{eqnarray}
     \langle \Delta z^2(n,\tau) \rangle \simeq
 \frac{\langle \Delta z_i^2(\tau) \rangle }{m(\tau)}
  \label{CF2}
 \end{eqnarray}
We refer to the previous equation as the compounding formula. This is based on the 
assumption that the tagged monomer MSD is governed independently by the dynamics of an 
isolated monomer which is modulated by the effect of the chain connectivity via the 
factor $m(\tau)$.

We note that the compounding formula, although not written down explicitly in the 
literature as in \eqref{CF2}, it is implicitly invoked in explaining the anomalous 
subdiffusive motion of  a tagged monomer in a Rouse chain in equilibrium 
$\langle \Delta z^2(n,\tau) \rangle_{eq} \sim \tau^{1/2}$. To explain this scaling 
it is argued that a tagged monomer drags along a growing portion of the polymer chain, 
which makes its effective friction increase with time. 
In \eqref{CF2}, the factor $m(\tau)$ is the growing effective friction, which 
as we shall argue scales as $m(\tau) \sim \tau^{1/2}$. The MSD of an isolated monomer 
follows an ordinary diffusive behavior $\langle \Delta z_i^2(\tau) \rangle \sim \tau$.
The ratio of these two factors gives indeed a tagged monomer MSD scaling as $\sim \tau^{1/2}$.

\subsection{Isolated monomer dynamics}
\label{sec:unconn}

We consider first the dynamics of an isolated monomer, unconnected from the rest 
of a polymer chain, which is governed by the following Langevin equation
\begin{eqnarray}
    \gamma \frac{\partial z_i(t)}{\partial t} = f(t)
\end{eqnarray}
where the random force $f(t)$ has an autocorrelation given by
\begin{eqnarray}
\langle f(t) f(t') \rangle = A g(|t-t'|)
\end{eqnarray}
Defining $\Delta z_i (\tau) \equiv z_i (\tau) - z_i(0)$, we find the mean-squared 
displacement
\begin{eqnarray}
    \langle \Delta z_i (\tau)^2 \rangle &=& 
    \frac{2A}{\gamma^2} \int_0^\tau dt' \int_0^{t'} dt'' \ g(t'-t'')  
    \nonumber \\
    &=& \frac{2A}{\gamma^2} \int_0^\tau du \ (\tau - u) \ g(u)
\label{supp:isolated}
\end{eqnarray}

For thermal noise $g(t)=\delta(t)$ and $A=2\gamma k_BT$ one finds from \eqref{supp:isolated}
the standard diffusive behavior 
\begin{eqnarray}
\langle \Delta z_i (\tau)^2 \rangle_{eq} = \frac{2 k_B T}{\gamma} \, \tau  
\end{eqnarray}
For noise of non-thermal origin, Eq.~\eqref{supp:isolated} typically predicts different 
scaling regimes. For instance, for an active Orstein-Uhlenbeck noise (AOU) 
$g(t) = e^{-t/\tau_A}$ one finds
\begin{eqnarray}
        \langle \Delta z_i (\tau)^2 \rangle &=& 
        \frac{2A\tau_A}{\gamma^2} \left[ \tau - \tau_A \left( 1 - e^{-\tau/\tau_A}\right) \right]
\end{eqnarray}
which gives
\begin{eqnarray}
    \langle \Delta z_i (\tau)^2 \rangle &=& \frac{A}{\gamma^2}
    \left\{
    \begin{array}{ccc}
        \tau^2 &  \quad &\tau \ll \tau_A\\
        2 \tau_A \tau  & \quad &\tau \gg \tau_A 
    \end{array}
    \right.
    \label{iso_mon}
\end{eqnarray}
At short times the noise is persistent $g(t) \approx 1$, which leads to a ballistic 
type of scaling behavior, while at long times one recovers the characteristic diffusive 
behavior. This crossover behavior occurs for other noises as well. For instance, 
a power-law correlated noise of the type
\begin{eqnarray}
    g(t) = \frac{1}{1 + (t/\tau_A)^\xi}
    \label{g_t_pow}
\end{eqnarray}
with $0 \leq \xi \leq 1$, Eq.~\eqref{supp:isolated} gives
\begin{eqnarray}
    \langle \Delta z_i (\tau)^2 \rangle &=& \frac{A}{\gamma^2}
    \left\{
    \begin{array}{ccc}
        \tau^2 &  \quad &\tau \ll \tau_A\\
        \tau_A^2 (\tau/\tau_A)^{2-\xi}  & \quad &\tau \gg \tau_A 
    \end{array}
    \right.
    \label{iso_mon_pow}
\end{eqnarray}

\subsection{Connectivity factor from tension propagation mechanism}
\label{sec_m}

Next we examine the factor $m(\tau)$ in the compounding formula \eqref{CF2}. 
The Rouse equation of motion~\eqref{eq:Rouse} takes a form of diffusion 
equation aside from the noise term. Neglecting boundary terms at the two 
polymer ends, the following Gaussian propagator
\begin{equation}
G(n,t) = \sqrt{\frac{\tau_0}{4 \pi t}} \ \exp \left( - \frac{\tau_0}{4t} n^2 \right)  
\label{def_prop}
\end{equation}
solves Eq.~\eqref{eq:Rouse}, where $\tau_0 = \gamma/k$ is the monomer time scale.
The propagator~\eqref{def_prop} implies that any local perturbation applied at 
monomer $n$ at time $t=0$ grows with a variance linear in time $\sigma^2(t) = 2 t/\tau_0$. 
The standard deviation gives then a measure of the number of monomers correlated to 
the perturbed monomer 
\begin{eqnarray}
m(\tau) \simeq (2 \tau/\tau_0)^{1/2},
\label{def_mtau_Rouse}
\end{eqnarray}
where $\tau=t-s$ is the time scale determined by the two measurement times 
$s$ and $t (>s)$. We call $m(\tau)$ the {\it connectivity factor}. Underlying 
the dynamics of polymers, we note that this scaling for the connectivity factor 
has been invoked, either implicitly or explicitly, in the description of various 
dynamical properties of polymers in equilibrium~\cite{Rubinstein_book}. It also plays a central role in the scaling description of polymer dynamics driven by external force as seen, for instance, in polymer translocation~\cite{Sakaue_2007, Grosberg_2011, Ikonen_2013,Fred14,Sakaue_2016,Sarabadani_2020,Micheletti_2017}.

For later use, it is useful to introduce the Fourier representation of the propagator;
\begin{eqnarray}
\widetilde{G}(q,t) \equiv \int_{-\infty}^{+\infty} dn \ G(n,t) \ e^{-iqn} 
= e^{-{t}/{\tau_q}}
\label{G_q_t}
\end{eqnarray}
with $\tau_{q} = \tau_0 q^{-2}$.
One can generalize this diffusive scaling to
\begin{eqnarray}
\tau_{q} =  \tau_0 q^{-\eta}
\label{tat_q}
\end{eqnarray}
where non-Gaussian exponent $\eta \neq 2$ has been invoked by various authors 
to describe complex phenomena not accounted for by the simple Rouse model.
For instance, values of $1 < \eta < 3$ describe Rouse models with distal 
couplings extending beyond neighboring monomers \cite{doi88,nool87,amit13,polo18,Saito_2015}, 
where $1 < \eta < 2$ or $2 < \eta < 3$ corresponds to effective attractive 
or repulsive interaction, respectively. As representative examples, a value 
$\eta=11/5$ or $\eta=5/3$  mimics the self-avoiding effect in three dimension 
or the crumpled globule conformation. In case hydrodynamic interactions are 
relevant, the friction becomes non-local, and $\eta = 3 \nu$ (with $\nu=3/5$ 
the Flory exponent) is used to describe Zimm dynamics in good solvent~\cite{doi88}.
For the generalized case \eqref{tat_q} one expects
\begin{eqnarray}
    m(\tau) \simeq (2\tau/\tau_0)^{1/\eta} ,
    \label{m_tau}
\end{eqnarray}
which represents a connectivity factor which is modified with respect
to that of the equilibrium Rouse model of Eq.~\eqref{def_mtau_Rouse}.

\begin{figure}[t]
    \centering
    \includegraphics[width=0.9\linewidth]{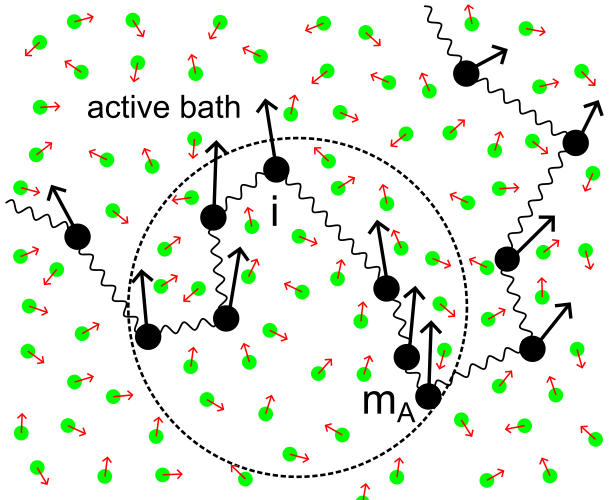}
    \caption{In an active bath at steady state $m_A$ monomers (Eq.~\eqref{m_A}) 
    are dynamically correlated, which leads to two different scaling relations 
    for transient and steady state dynamics, Eqs.~\eqref{MSD_CF_prediction_tr} 
    and \eqref{MSD_CF_prediction_ss}. This correlation can be quantified from 
    the calculation of displacement correlations reported in Sec.~\ref{sec:H}, 
    showing that all monomers in the block $m_A$ perform an averaged correlated 
    motion (black arrow).}
    \label{fig:Rouse_distal}
\end{figure}

We now aim to extend the scope of the connectivity factor by applying to the 
active polymers. In doing so, we need to take account of one new aspect arising 
from the noise persistence. Noises like the AOU one $g(t)=e^{-t/\tau_A}$ or 
power-law correlated noises as \eqref{g_t_pow} are characterized by a persistence 
time $\tau_A$, or to an associated number of monomers
\begin{eqnarray}
m_A \equiv \left(\frac{2 \tau_A}{\tau_0} \right)^{1/\eta}
\label{m_A}
\end{eqnarray}
In other words, the active noise with persistence time $\tau_A$ induces a
steady state dynamical correlation between $m_A$ monomers, as illustrated
in Fig.~\ref{fig:Rouse_distal}. This dynamical correlation can be best
understood from the calculation of displacement correlations presented
in Sec.~\ref{sec:H}. This leads us to distinguish two protocols to perform 
the measurement:
\begin{itemize}
    \item[(i)]  In {\it transient} case, the polymer is settled in thermal equilibrium 
    and active noise is switched on at time $s$. Physical quantities probed at $t (>s)$ 
    reflect the transient process towards a new steady state. 
     \item[(ii)]  In {\it steady state} case, active noise is switched on in distant 
     past, at time $-T_{\infty}$, with $T_{\infty}$ much longer than any other 
     characteristic relaxation times of the system. Here one probes properties of the 
     system in its active steady state. 
\end{itemize}
To summarize, the tension propagation mechanism in polymer chain implies the growing 
connectivity factor with time scale. However, in the case (ii) above, the active noise 
with the persistence time $\tau_A$ applied in the time interval $[-T_{\infty}, s]$ 
generates an active steady state consisting of correlated domains of $m_A$ monomers.
This leads to the following scaling formula for the connectivity factor;
\begin{eqnarray}
m(\tau) \simeq
    \left\{
    \begin{array}{ccc}
       \left(\frac{2 \tau}{\tau_0} \right)^{1/\eta} &  ({\rm transient})\\
        m_A+\left(\frac{2 \tau}{\tau_0} \right)^{1/\eta}  &({\rm steady \ state}) 
    \end{array}
    \right.
\label{m_tau_general}
\end{eqnarray}
In the white noise limit $\tau_A \rightarrow 0$ (thus $m_A \rightarrow 0$, 
Eq.~\eqref{m_A}), the two expressions become identical as expected. In Sec.~\ref{sec:H} 
we present a rigorous calculation of $m_A$ for the diffusive case $\eta=2$.

\subsection{Scaling predictions}
\label{scaling_prediction}

Combining the connectivity factor~\eqref{m_tau_general} and the isolated monomer dynamics~\eqref{iso_mon},
the compounding formula~\eqref{CF2} yields the distinct (i) transient and (ii) steady state 
scaling formulae for the MSD of tagged monomer in active polymer. Omitting numerical coeffcients 
of order unity, these are respectively given as
\begin{itemize}
\item transient
\begin{eqnarray}
\langle \Delta z^2(n,\tau) \rangle_{tr}
\sim
\left\{
\begin{array}{ll} 
   \frac{A}{\gamma^2} \tau_0^{2} \left(\frac{\tau}{\tau_0} \right)^{2-1/\eta}  &  ( \tau \ll \tau_A) \\ 
  \frac{A}{\gamma^2}  \tau_0 \tau_A \left(\frac{\tau}{\tau_0} \right)^{1- 1/\eta}  &  ( \tau \gg \tau_A)
\end{array}
\right.
\label{MSD_CF_prediction_tr}
\end{eqnarray}

\item steady state
\begin{eqnarray}
\langle \Delta z^2(n,\tau) \rangle_{ss}
\sim
\left\{
\begin{array}{ll} 
   \frac{A}{\gamma^2}\tau_0^2  (\frac{\tau_0}{\tau_A})^{1/\eta}  
   \left( \frac{\tau}{\tau_0} \right)^{2}  &  ( \tau \ll \tau_A) \\ 
  \frac{A}{\gamma^2}  \tau_0 \tau_A \left(\frac{\tau}{\tau_0} \right)^{1- 1/\eta}  &  ( \tau \gg \tau_A)
\end{array}
\right.
\label{MSD_CF_prediction_ss}
\end{eqnarray}
\end{itemize}
We observe that the two cases above share the same long time scale 
scaling $(\tau \gg \tau_A)$. Interestingly, the short time steady 
state behavior $\sim \tau^2$ is super-universal as it is independent 
from the value of $\eta$. Here the type of polymer connectivity (e.g. 
ideal $\eta=2$ or distal long range $\eta \neq 2$) does not influence 
the short time behavior. A tagged monomer performs a correlated motion 
with a constant number ($m_A$) of surrounding monomers. The exponent 
$\eta$ influences the value of $m_A$, hence the prefactor, but not the 
time scaling. Notice that if the noise has a long persistence given by 
Eq.~\eqref{g_t_pow}, the long time scale scaling is modified, and 
given by the following formula (valid for $\tau \gg \tau_A$)
\begin{eqnarray}
\langle \Delta z^2(n,\tau) \rangle_{tr, ss}
&\sim& 
  \frac{A}{\gamma^2} \tau_0^2  \left( \frac{\tau_A}{\tau_0} \right)^{\xi} 
  \left(\frac{\tau}{\tau_0} \right)^{2-\xi -1/\eta}  
  \label{MSD_CF_prediction_pw_noise}
\end{eqnarray}
again for both transient and steady state cases.

\section{Rigorous approach}
\label{sec:rigorous_approach}

We now provide a systematic analysis starting from the Rouse equation of 
motion~\eqref{eq:Rouse} to verify the scaling predictions from the compounding 
formula approach. We take two approaches that are complementary to each other. 
The first is the normal mode analysis (Sec.~\ref{sec:normal_mode_analysis}), 
which allows us to generalize the Rouse model to include the effect of distal 
couplings through the introduction of an exponent $\eta \neq 2$ via the 
generalized scaling of Eq.~\eqref{tat_q}. The result of this analysis are 
the MSD expressions for the steady state \eqref{MSD_ss_A} and for the transient 
case \eqref{MSD_tr_A}. This extends the analysis of Ref.~\cite{osma17}, which 
considered the steady-state case for $\eta=2$. In the second approach 
(Sec.~\ref{sec:real_space_analysis}), we solve the equation~\eqref{eq:Rouse} 
directly in the real space using the propagator~\eqref{def_prop}. While 
restricted to the Rouse model ($\eta =2$), the real space approach provides 
an exact yet compact MSD formula for general noise correlator $g(t)$, 
distinguishing again between steady state \eqref{app:res_ss} and transient 
\eqref{app:res_tr} cases. In addition, it also enables us to obtain a concise 
expression for the displacement correlation functions (see Sec.~\ref{sec:H}) 
that allows us to identify the connectivity factor $m(\tau)$ in an 
unambiguous way.

\subsection{Normal mode analysis}
\label{sec:normal_mode_analysis}

Given $z(n,t)$ solution of Eq.~\eqref{eq:Rouse}, we introduce the transformation 
to the normal mode
\begin{eqnarray}
Z_p(t) \equiv \frac{1}{N}\int_0^N dn \  z(n,t) \cos{\left(\frac{p \pi n}{N}\right)}
\end{eqnarray}
for $p=0,1,2, \cdots$. Its inverse transform is
\begin{eqnarray}
z(n,t) = Z_0(t) + 2 \sum_{p=1} Z_p(t) \cos{\left(\frac{p \pi n}{N}\right)}
\end{eqnarray}
In the continuum description, the summation over $p$ is formally taken 
to $p=\infty$. In reality, however, the discrete nature of the polymer 
introduces the upper cut-off $p_{max}=N$, which corresponds to the short 
wave length cut-off at the monomer scale. The Rouse equation of motion
(\ref{eq:Rouse}), for normal modes, takes the form
\begin{eqnarray}
\gamma \frac{d Z_p(t)}{dt} = -k_p Z_p(t) + F_p(t)
\label{eq:Rouse_n}
\end{eqnarray}
where 
\begin{eqnarray}
\langle F_p(t) \rangle &=& 0  \label{F_p_ave1} \\
\langle F_p(t) F_q(t') \rangle &=& \frac{1 + \delta_{p0}}{2N} A \, g(|t-t'|) \delta_{pq}
 \label{F_p_ave2}
\end{eqnarray}
with $g(u)$ the noise correlator, as given in Eq.~\eqref{auto_corr_noise}.
Note that the independence of $F_p(t)$ among different modes arises from 
the assumption that independent noise $f(n,t)$ is acting on each monomer.
The spring constant $k_p = k (\pi p/N)^2$ for the mode $p$ defines the 
relaxation time $\tau_p = \gamma/k_p = \tau_0 (N/p \pi)^2$. In the 
following, we generalize these relations as
\begin{eqnarray}
k_p &=& k (\pi p/N)^{\eta} \\
\tau_p &=& \gamma/k_p = \tau_0 (N/p \pi)^{\eta} = \frac{\tau_R}{p^{\eta}} \label{tau_p}
\end{eqnarray}
where $\tau_R \equiv \tau_1$ is the longest relaxation time. With the wave number 
$q=p \pi/N$, Eq.~\eqref{tau_p} corresponds to Eq.~\eqref{tat_q}. 

Equation~(\ref{eq:Rouse_n}) is solved as
\begin{eqnarray}
Z_p(t) = Z_p(t_0) e^{-(t-t_0)/\tau_p} + 
\frac{1}{\gamma} \! \int_{t_0}^t \!e^{-(t-t')/\tau_p} F_p(t') dt' 
\nonumber \\
\label{sol:Rouse_n}
\end{eqnarray}
where $Z_p(t_0)$ is the initial condition at $t=t_0$.
The MSD of the tagged monomers in the time interval $t-s$ is calculated as
\begin{eqnarray}
\langle (z(n,t) - z(n,s))^2 \rangle = \langle (Z_0(t) - Z_0(s))^2 \rangle 
\nonumber\\
+ 4 \sum_{p \ge 1} \langle (Z_p(t) - Z_p(s))^2 \rangle \cos^2{\left( \frac{p \pi n}{N}\right)}
\label{MSD_mode_1}
\end{eqnarray}
where we have used the independence of modes. For long polymers, the longest relaxation 
time $\tau_R \equiv \tau_1$ is very long, and in the time scale $\tau \ll \tau_R$, the 
first term in Eq.~(\ref{MSD_mode_1}), i.e., the center-of-mass mode is negligible. 
In addition, except for the monomers close to chain ends, we can replace the square 
cosine term as $\cos^2{\left( \frac{p \pi n}{N}\right)} = 1/2$. In the following, we 
thus focus on the following expression;
\begin{eqnarray}
\langle (z(n,t) - z(n,s))^2 \rangle = 2 \sum_{p \ge 1} \langle (Z_p(t) - Z_p(s))^2 \rangle 
\label{MSD_mode_2}
\end{eqnarray}

We now calculate the time-correlation function $C_p(t,s;t_0) \equiv \langle 
Z_p(t) Z_p(s) \rangle_{t_0}$ of the mode $p$. Since this quantity, in general, 
depends on the initial condition, we explicitly include the initial time $t_0$ 
in the subscript. From the solution~(\ref{sol:Rouse_n}), and the properties 
of random forces \eqref{F_p_ave1} and \eqref{F_p_ave2} we obtain
\begin{eqnarray}
&&C_p(t,s;t_0) = \langle Z_p(t_0)^2 \rangle 
e^{-(t+s-2t_0)/\tau_p}
\nonumber \\
&&+ \frac{A}{2 N \gamma^2}\int_{t_0}^{t} dt' \int_{t_0}^{s} ds' 
e^{-(t+s-t'-s')/\tau_p} g(|t' - s'|) 
\nonumber \\
\label{C_p_t0}
\end{eqnarray}
Here we introduce a distinction between transient and steady state. If the 
system is sufficiently aged, i.e., $t_0 \rightarrow -\infty$, it would be 
settled in its steady-state, in which the time-translational invariance 
implies that the correlation function depends only on the time difference 
$\tau = |t-s|$. We thus define the steady-state correlation function by 
letting $t_0 \rightarrow -\infty$
\begin{eqnarray}
C^{(ss)}_p(\tau) &\equiv& C_p(s+\tau,s;-\infty)
\label{def:Css}
\end{eqnarray} 
Note that in this limit the first term on the right hand side of
Eq.~\eqref{C_p_t0} vanishes and, as expected, the MSD becomes 
independent on the initial condition $\langle Z_p(t_0)^2 \rangle$.
As we have $\langle Z_p^2(t)\rangle_{ss} = \langle Z_p^2(s)\rangle_{ss} 
= C^{(ss)}_p(0)$ and $\langle Z_p(t) Z_p(s)\rangle_{ss} = 
C^{(ss)}_p(\tau)$, the steady-state MSD is given by
\begin{eqnarray}
\langle \Delta z^2(n,\tau) \rangle_{ss} &=& 
4 \sum_{p \ge 1}  (C^{(ss)}_p(0)  - C^{(ss)}_p(\tau) ) 
\label{MSD_mode_ss}
\end{eqnarray}
Another case of interest is the transient process, where we set $t_0=s=0$ and 
$\tau = t (=t-s)$. In this case, the measurement starts immediately after the 
system is prepared at time $t_0$. The transient MSD $\langle \Delta z^2(n,\tau) 
\rangle_{tr}  \equiv \langle (z(n,t) - z(n,s))^2 \rangle_{tr}$ is represented as
\begin{eqnarray}
\langle \Delta z^2(n,\tau) \rangle_{tr}  \equiv \langle (z(n,t) - z(n,s))^2 \rangle_{tr}
\nonumber\\ 
= 2 \sum_{p \ge 1}  [C_p(t,t;0) + C_p(0,0;0)   - 2C_p(t,0;0) ]
\label{MSD_mode_tr}
\end{eqnarray}

To proceed further, one needs to specify the noise $f(n, t)$. In all generality, 
we assume that the noise is a sum of thermal and active contributions $f(n,t) =
f_{th}(n,t) + f_a(n,t)$. Each noise has the auto-correlation of the form 
\eqref{auto_corr_noise} or \eqref{F_p_ave2}, where the relations \eqref{FDT} hold 
for the thermal noise $f_{th}$. Since these two noise sources are independent, their 
contributions can be calculated separately, and the net MSD is obtained as their 
addition. The calculation of the thermal contribution is found in a standard 
textbook \cite{doi88}, which we outline in Appendix~\ref{sec:equil-final}. 
Here we focus on the contribution of the active noise, for which we assume 
the AOU form of the correlator $g(t) = e^{-t/\tau_A}$. The following calculation 
generalizes the results of Ref.~\cite{osma17} by considering the transient as well 
as the steady state scenarios and extends the results to a generalized Rouse model 
characterized by an exponent $\eta \neq 2$, see also \cite{Gompper_2020}.
With $t>s$, we obtain
\begin{eqnarray}
&&C_p(t,s;t_0) = \langle Z_p(t_0)^2 \rangle e^{-(t+s-2t_0)/\tau_p} 
\nonumber \\
&&+ \frac{A}{2N \gamma^2} 
\frac{\tau_A \tau_p^2}{\tau_A^2-\tau_p^2}
\left(\tau_A e^{-\tau/\tau_A} - \tau_p e^{-\tau/\tau_p}
\right)
\nonumber\\
&&+ \frac{A}{2N \gamma^2} \, h_p(t,s;t_0)
  \label{C_p_A-new}
\end{eqnarray}
with $\tau \equiv t-s$ and
\begin{eqnarray}
&&    h_p(t,s;t_0) \equiv
    \frac{\tau_p^2 \tau_A}{\tau_A -\tau_p} e^{-(t+s-2t_0)/\tau_p}
    \nonumber \\
    &&- \frac{\tau_p^2 \tau_A^2}{\tau_A^2 -\tau_p^2} 
    \left[
    e^{-(t-t_0)/\tau_p} e^{-(s-t_0)/\tau_A} 
    \right.
    \nonumber \\
    && + 
    \left.
    e^{-(s-t_0)/\tau_p} e^{-(t-t_0)/\tau_A}
    \right]
\end{eqnarray}

\paragraph{Steady state}
By letting $t_0 \rightarrow -\infty$, we obtain the steady-state correlation 
function (see \eqref{def:Css}). Note that the $t_0$-dependent terms on the 
left hand side of Eq.~\eqref{C_p_A-new} vanish, so one gets
\begin{eqnarray}
C^{(ss)}_p(\tau) = \frac{A}{2N \gamma^2} 
\frac{\tau_A \tau_p^2}{\tau_A^2-\tau_p^2}
\left(\tau_A e^{-\tau/\tau_A} - \tau_p e^{-\tau/\tau_p}
\right)
\nonumber \\
\label{C_p_A_ss}
\end{eqnarray}
which only depends on the time difference $\tau= t-s$. Using Eqs.~(\ref{MSD_mode_ss}) 
and (\ref{C_p_A_ss}), the steady-state MSD is obtained as
\begin{eqnarray}
&& \langle \Delta z^2(n,\tau) \rangle_{ss} = 
\frac{2 A  \tau_A }{N \gamma^2 } \sum_{p =1}^N \frac{\tau_p^2}{\tau_A^2 - \tau_p^2}
\nonumber\\
&& 
\times \left[
\tau_A \left(1-e^{-\tau/\tau_A} \right) -
\tau_p \left(1-e^{-\tau/\tau_p} \right) 
\right]
\label{MSD_ss_A}
\end{eqnarray}

\paragraph{Transient}
To calculate the transient MSD from Eq.~\eqref{MSD_mode_tr}, we first  
evaluate the following correlation functions:
\begin{eqnarray}
    &&C_p(t,t;0) = \langle Z_p(0)^2\rangle e^{-2t/\tau_p} 
+\frac{A}{2 N \gamma^2} \frac{\tau_A\tau_p^2}{\tau_A^2-\tau_p^2}
\nonumber \\
&&\times \!\!
\left[  
\tau_A - \tau_p + (\tau_A+\tau_p) e^{-2t/\tau_p}-2 \tau_A e^{-t(1/\tau_p+1/\tau_A)}
\right] 
\nonumber \\
\\
&& C_p(0,0;0) = \langle Z_p(0)^2\rangle \\ 
&& C_p(t,0;0) = \langle Z_p(0)^2\rangle e^{-t/\tau_p} 
\end{eqnarray}
In general, the transient MSD depends on the initial configuration $Z_p(0)$. In 
the transient protocol defined in Sec.~\ref{sec_m}, the term $\langle Z_p(0)^2\rangle$ 
obeys the equipartition theorem, as the polymer is in thermal equilibrium 
at $t=0$, as pure thermal noise is present in the interval $[-\infty,0]$.
From  Eq.~\eqref{MSD_mode_tr} and the above correlation functions, and noting 
$\tau=t$ in our definition of the transient process, we obtain the transient 
MSD as
\begin{eqnarray}
&&\langle \Delta z^2(n,\tau) \rangle_{tr} =  2 \sum_{p=1}^N  
\left\{ \langle Z^2_p(0) \rangle    (1-e^{-\tau/\tau_p})^2 
\right.
\nonumber \\
&&+ \frac{A}{2 N \gamma^2}  \frac{\tau_A\tau_p^2}{\tau_A^2-\tau_p^2}
\left[  
\tau_A - \tau_p + (\tau_A+\tau_p) e^{-2\tau/\tau_p}
\right.
\nonumber\\
&&\left.
\left.
-2 \tau_A e^{-\tau(1/\tau_p+1/\tau_A)}
\right]
\right\}
\label{MSD_tr_A}
\end{eqnarray}
Note that \eqref{MSD_ss_A} contains only the active contribution to the steady 
state MSD, while in the transient MSD the thermal noise act only up to time $t=0$.
Quantitatively, the actual MSD in each case is supplemented by the thermal 
contribution over the full time interval, which will be further commented on 
the subsequent sections.

\begin{figure}[t]
    \centering
    \includegraphics[width=1\linewidth]{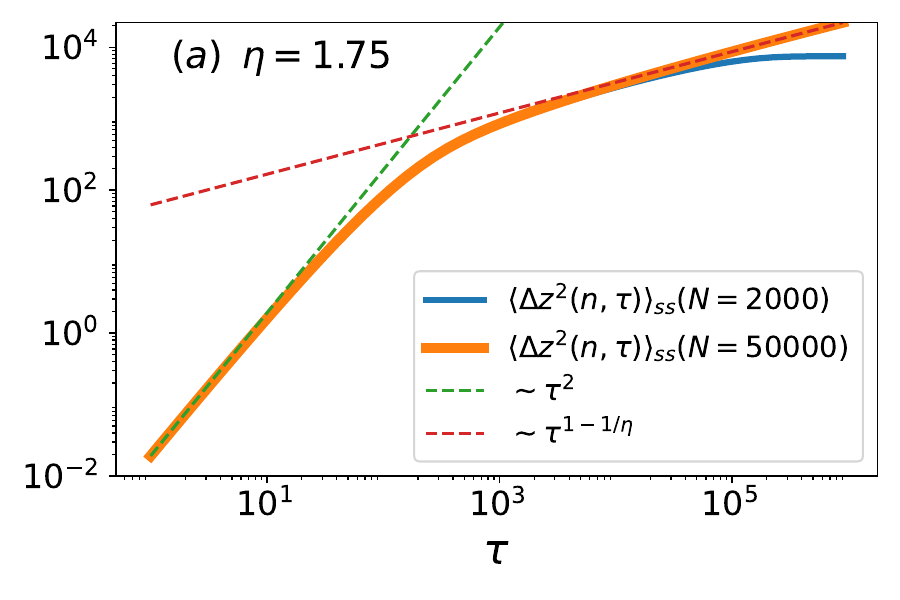}
    \includegraphics[width=1\linewidth]{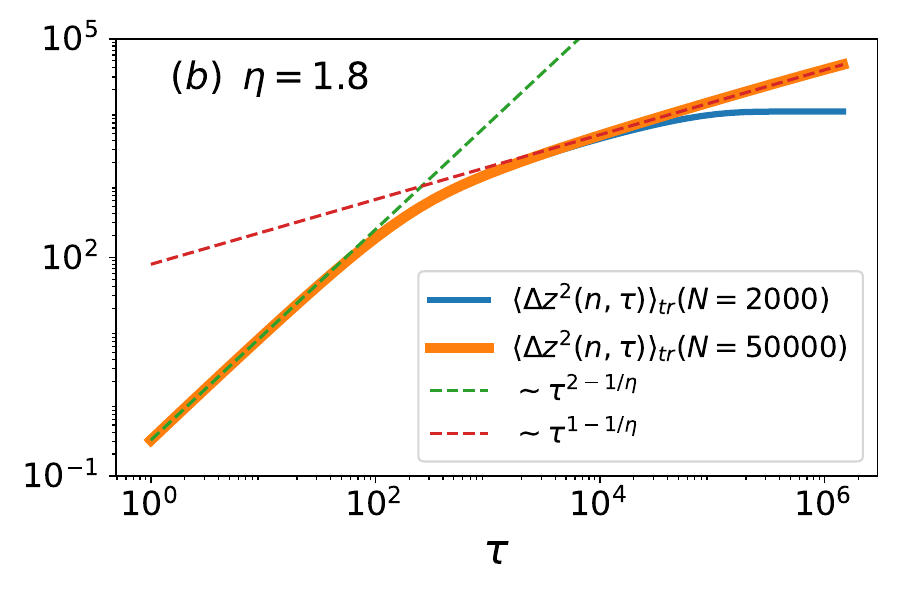}
    \caption{Normal mode analysis for tagged monomer MSD for an active
    Ornstein-Uhlenbeck noise. (a) Steady-state MSD (Eq.~\eqref{MSD_ss_A}) for 
    the generalized Rouse model with $\eta=1.75$. (b) Transient MSD 
    (Eq.~\eqref{MSD_tr_A}) for $\eta=1.8$. The parameter $\tau_A=100$ and 
    $\tau_0=1$ were used. The sums are extended to two $N$ (number of normal 
    modes) corresponding to polymers of different lengths. The MSD saturates 
    at the Rouse time $\tau_R$ as Eqs.~\eqref{MSD_ss_A} and \eqref{MSD_tr_A} do 
    not take into account the center of mass motion. As $\tau_R \sim N^{\eta}$ 
    the saturation is visible for the two shorter $N$. The dashed lines are the 
    predictions of the compounding formula approach, Eqs.~\eqref{MSD_CF_prediction_tr} 
    and \eqref{MSD_CF_prediction_ss}.}
    \label{fig:Rouse_gen}
\end{figure}

Figure \ref{fig:Rouse_gen} shows the steady state and transient MSDs numerically 
evaluated from Eqs~\eqref{MSD_ss_A} and~\eqref{MSD_tr_A}, respectively. In the transient
case we dropped the term proportional to $\langle Z^2_p(0)\rangle$, assuming a dominant
active noise contribution ($A$ large). We use exponents $\eta =1.75$ and $\eta=1.8$ 
(see Eq.~\eqref{tau_p}) deviating from the ordinary Rouse model $\eta=2$. Both plots 
show two scaling regimes with distinct power-law behaviors for $\tau \ll \tau_A$
and $\tau \gg \tau_A$ (in the plots $\tau_A = 100$). The behavior is in agreement with 
the predictions from the compounding formula approach given by \eqref{MSD_CF_prediction_tr}
and \eqref{MSD_CF_prediction_ss}.

\subsection{Real space analysis}
\label{sec:real_space_analysis}

We now solve the Rouse equation of motion~\eqref{eq:Rouse} in real space 
using the propagator~\eqref{def_prop}. The advantage of this approach
is that it leads to more compact expressions for the MSD, valid for any type 
of noise, although limited to the ordinary Rouse model, corresponding to $\eta=2$
in the previous normal mode analysis. The end results of this analysis are the 
MSD expressions for the steady state and transient cases, which are given by 
Eqs.~\eqref{app:res_ss} and \eqref{app:res_tr}, respectively. The approach can be 
generalized to the calculation of the correlation displacement in Sec.~\ref{sec:H}.

The solution of Eq.~\eqref{eq:Rouse} at time $t$, given an initial condition 
$z(n,s)$ at some earlier time $s <t$ can be written as
\begin{eqnarray}
    z(n,t;s) &=& \frac{1}{\gamma}  \int dn' \int_s^t dt' \ G(n-n',t-t') \ f(n',t') 
    \nonumber \\
    &+& 
    \int dn' \ G(n-n',t-s) \ z(n',s)
\label{gen_sol}
\end{eqnarray}

The integration in $n'$ in \eqref{gen_sol} is extended over the whole real domain, 
corresponding to an infinitely long polymer. In the limit $t \to s$ the first 
integral in \eqref{gen_sol} vanishes and the second reduces to $z(n,s)$ as 
$\lim_{t \to s} G(n-n',t-s) = \delta (n-n')$. We are interested in the calculation 
of the displacement of the $n$-th monomer over a time interval
$\tau \equiv t-s$, which is given by
\begin{eqnarray}
 &&   \Delta z_n (s \to t; -T_\infty) \equiv z(n,t;-T_\infty) -  z(n,s;-T_\infty)
    \nonumber \\
 &&   = \frac{1}{\gamma} \int dn' \left[
    \int_{-T_\infty}^t dt' \ G(n-n',t-t') \ f(n',t')   
    \right.
    \nonumber\\
&&    -
    \left.
    \int_{-T_\infty}^s dt' \ G(n-n',s-t') \ f(n',t')
    \right]
\label{Dz1}
\end{eqnarray}
where we consider $T_\infty$ sufficiently large, hence, allowing us to omit the dependence on the initial configuration $z(n,-T_\infty)$.
Adding and subtracting to \eqref{Dz1} the quantity
\begin{equation}
    \frac{1}{\gamma} \int dn'  \int_{-T_\infty}^s dt' \ G(n-n',t-t') \ f(n',t')
\end{equation}
we can rewrite the displacement as
\begin{eqnarray}
    \Delta z_n (s \to t; -T_\infty) &=& {\cal A}_n(\tau) + {\cal B}_n(\tau)
    \label{dz-decomposition}
\end{eqnarray}
with
\begin{eqnarray}
    {\cal A}_n(\tau) &\equiv& \frac{1}{\gamma} \! \int dn' \! \int_{s}^t \! \! dt' \ G(n-n',t-t') \ f(n',t')
     \label{defA} \nonumber \\
     \\ 
    {\cal B}_n(\tau) &\equiv& \frac{1}{\gamma} \! \int dn' \! \int_{-T_\infty}^s \!\! dt' \left[ G(n-n',t-t') 
    \right.
    \nonumber \\
    &-& \left. G(n-n',s-t')\right]\ f(n',t')
    \label{defB}
\end{eqnarray}
where $\tau=t-s$.
The mean squared displacement is given by
\begin{eqnarray}
    \langle \Delta z_n^2 \rangle  (s \to t; -T_\infty) &=& 
    \langle {\cal A}_n (\tau) ^2\rangle + \langle {\cal B}_n(\tau) ^2\rangle \nonumber\\
    &+& 2\langle {\cal A}_n(\tau) {\cal B}_n(\tau) \rangle 
    \label{MSD_A-B}
\end{eqnarray}
where the average $\langle . \rangle$ is performed over the noise realizations.
We find in the limit $T_\infty \to \infty$ (the details of the calculations are given 
in Appendix~\ref{subsec:details})
\begin{widetext}
\begin{eqnarray}
   \langle {\cal A}_n(\tau)^2 \rangle &=& \frac{A}{\gamma^2} \sqrt{\frac{\tau_0}{\pi}} 
            \int_0^\tau du \ g(u) \left[ \sqrt{2\tau -u} - \sqrt{u} \right]
            \label{resA2}
    \\
    \langle {\cal B}_n(\tau)^2\rangle &=& \frac{A}{\gamma^2}  \sqrt{\frac{\tau_0}{\pi}} 
            \int_0^{+\infty} \!\!\!\!\!\! du \ g(u) 
            \left[ 2 \sqrt{u+\tau} - \sqrt{u+2\tau} - \sqrt{u} \right]
            \label{resB2}
    \\
    2 \langle {\cal A }_n(\tau) {\cal B}_n (\tau) \rangle &=& \frac{A}{\gamma^2}  \sqrt{\frac{\tau_0}{\pi}} 
    \left[
    \int_0^\tau du\ g(u) \left( \sqrt{2\tau+u} - \sqrt{2\tau-u} - 
    \sqrt{\tau+u} + \sqrt{\tau-u} \right) + \right. \nonumber \\
    &&      \left.      \int_\tau^{+\infty} \!\!\!\!\!\! du \ g(u) 
    \left( \sqrt{2\tau+u} - \sqrt{u} - \sqrt{\tau+u} + \sqrt{u-\tau} \right)
    \right]
            \label{resAB}
\end{eqnarray}
where the noise correlation~\eqref{auto_corr_noise} has been used.
\end{widetext}

We first present the result for the steady state, then, proceed to the transient. 
In both cases, we focus on the active contribution, while the thermal contribution 
is presented in Appendix \ref{sec:equil-final}.

\paragraph{Steady state}
Recalling the noise consists of two independent sources 
\begin{eqnarray}
    f(n,t)=f_{th}(n,t) + f_a(n,t), 
    \label{f_ss}
\end{eqnarray}
the MSD is represented as
\begin{eqnarray}
   && \langle \Delta z_n^2 ( \tau) \rangle_{ss} =  
   \langle {\cal A}_n(\tau)^2\rangle_a +   
   2\langle {\cal A}_n(\tau) {\cal B}_n(\tau)\rangle_a 
   \nonumber \\
   && + \langle {\cal B}_n(\tau)^2\rangle_a + 
    \langle {\cal A}_n(\tau)^2 \rangle_{th} +
    \langle {\cal B}_n(\tau)^2 \rangle_{th}
    \end{eqnarray}
where the subscript (th or a) denotes the noise source. Note the 
absence of the cross term $\langle {\cal AB}\rangle$ for the thermal 
contribution (see Appendix) due to its white noise nature. Assuming 
the dominance of the active contribution, we sum up the three terms 
\eqref{resA2}, \eqref{resB2} and \eqref{resAB} and find
\begin{eqnarray}
 \langle \Delta z_n^2 (\tau)\rangle_{ss} =
    \langle {\cal A}_n(\tau)^2\rangle + \langle {\cal B}_n(\tau)^2\rangle + 
    2\langle {\cal A}_n(\tau) {\cal B}_n(\tau)\rangle \nonumber \\
\!\!\! = \frac{A}{\gamma^2}  \sqrt{\frac{\tau_0}{\pi}} \!\!
    \int_0^{+\infty} \!\!\!\!\!\!\!\!
    du  g(u) \!\! \left( \sqrt{\tau+u} - 2 \sqrt{u} + \sqrt{|\tau-u|} \right) 
    \nonumber\\
    \label{app:res_ss}
\end{eqnarray}

\begin{figure}[!t]
    \centering
    \includegraphics[width=1\linewidth]{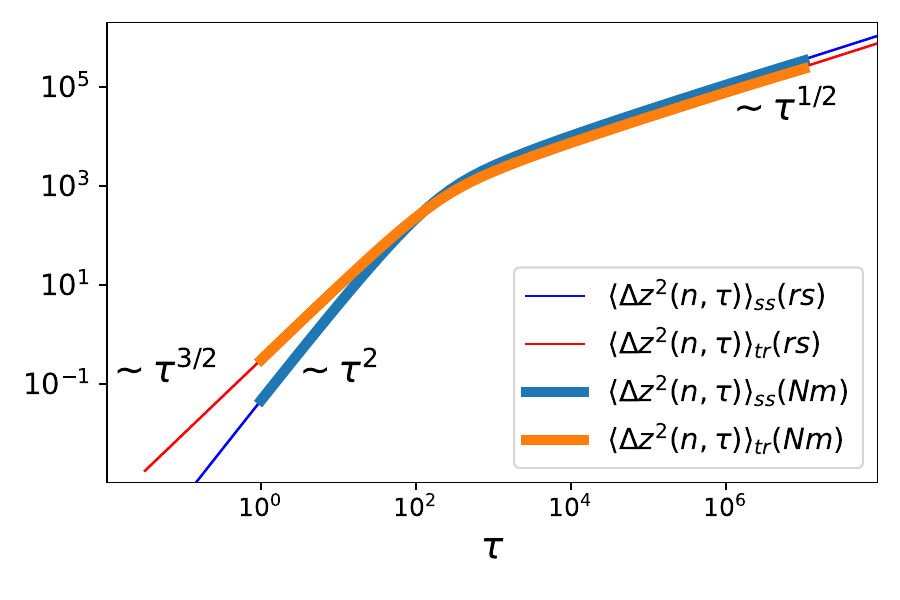}
    \caption{Comparison between the MSD for the real space (thin lines) 
    and normal mode (thick lines) for the AOU noise, showing the distinct 
    scaling regimes for the steady state and transient cases.}
    \label{fig:compare}
\end{figure}

\paragraph{Transient}
In the transient case we apply equilibrium dynamics during the whole 
process. At time $t>s$, an additional active noise term is turned on. 
The stochastic noise is therefore
\begin{eqnarray}
    f(n,t) = f_{th}(n,t) + \theta(t-s) f_a(n,t)
    \label{f_tr}
\end{eqnarray}
with $\theta(t)$ the Heaviside step function. From their definitions, 
see Eqs.~\eqref{defA} and~\eqref{defB}, ${\cal A}_n$ has both thermal 
and active components, while ${\cal B}_n$ has only thermal component. 
The MSD in this transient case is given by 
\begin{eqnarray}
    \langle \Delta z_n^2 ( \tau) \rangle_{tr} &=& 
    \langle {\cal A}_n(\tau)^2 \rangle_a + \langle {\cal A}_n(\tau)^2 \rangle_{th} +
    \langle {\cal B}_n(\tau)^2 \rangle_{th} \nonumber \\
\label{tr_AB}
\end{eqnarray}
Again, assuming the dominance of the active contribution, we find
\begin{eqnarray}
    \langle \Delta z_n^2 ( \tau) \rangle_{t r}&=& 
    \langle {\cal A}_n(\tau)^2 \rangle_a
    \nonumber\\
    & = &  
    \frac{A}{\gamma^2} \sqrt{\frac{\tau_0}{\pi}} \!\!
      \int_0^\tau \!\! du \ g(u) \left[ \sqrt{2\tau -u} - \sqrt{u} \right]
    \nonumber \\
\label{app:res_tr}
\end{eqnarray}
Figure~\ref{fig:compare} shows a comparison between the MSD obtained 
from the real-space and normal mode (with $\eta=2$) analysis.
To summarize, the above real space analysis provides a transparent 
distinction between steady state and transient scenarios. A comparison 
of Eq.~\eqref{app:res_ss} with Eq.~\eqref{app:res_tr} points to the role 
of ${\cal B}_n(\tau)$ in the steady state. We will elucidate in 
Sec.~\eqref{conc:interpret} that the term ${\cal B}_n(\tau)$ reflects 
the relaxation of the conformational degrees of freedom. As shown in 
Appendix \ref{sec:equil-final}, the thermal contribution is 
$\langle {\cal A}_n(\tau)^2 \rangle_{th}  \sim 
\langle {\cal B}_n(\tau)^2 \rangle_{th}  \sim \tau^{1/2}$, the effect 
of which may be apparent in the short and also in long time scales 
depending on the magnitude and the persistence of the active noise.
Note the absence of the cross term $\langle {\cal A}_n {\cal B}_n \rangle$ 
in the transient set-up (Eq.~\eqref{tr_AB}), where ${\cal B}_n(\tau)$ 
has only the thermal component. This cross-term is non-vanishing only for 
colored noise. From Eqs.~\eqref{defA} and~\eqref{defB} it follows that
$\delta$-correlated random forces $f(n,t)$ lead to $\langle {\cal A}_n {\cal B}_n 
\rangle =0$ as ${\cal A}_n$ and ${\cal B}_n$ are expressed as integrals in two
disjoint time domains.

\section{Displacement correlation}
\label{sec:H}

The correlation in displacement of two monomers $n$ apart along the chain during 
the time scale $\tau$ can be quantified by the following correlation function;
\begin{equation}
H(n, s \rightarrow t; t_0) \equiv \langle \Delta z_{n_1}(s \rightarrow t; t_0) 
\Delta z_{n_2}(s \rightarrow t; t_0) \rangle
\end{equation}
where $n= |n_1 - n_2|$. We note that by setting $n=0$, the above 
reduces to the tagged monomer MSD. As for the MSD, also 
the displacement correlation can be defined for steady state and transient 
cases and its calculation follows quite closely the real space analysis of 
Sec.~\ref{sec:real_space_analysis}. The details of these calculations are reported 
in Appendix~\ref{app:displacement}. We find for the steady state
\begin{eqnarray}
H_{ss}(n,\tau) &=& \frac{A}{\gamma^2}
 \int_0^{+\infty} \!\!
du \, g(u) \, [ F_n(\tau+u)   \nonumber \\
&& - 2 F_n(u) + F_n(|\tau-u|)  ]
\label{Hss}
\end{eqnarray}
where we have defined 
\begin{eqnarray}
    F_n (\tau) &\equiv& \int_0^\tau dw \,  G(n,w)
    \nonumber \\
    &=& \frac{n \tau_0}{2} \left\{ 
    \frac{e^{-\displaystyle{\alpha^2(\tau)}}}{ \sqrt{\pi} \alpha(\tau)}
    - \text{Erfc}[\alpha(\tau)] \right\}
\label{def_Fn}
\end{eqnarray}
with $G(n,w)$ the Gaussian propagator \eqref{def_prop}, $\alpha(\tau) \equiv 
\frac{n}{2} \sqrt{\frac{\tau_0}{\tau}}$. $\text{Erfc(x)}$ is the complementary 
error function.

In the transient case, we find (see Appendix~\ref{app:displacement})
\begin{eqnarray}
H_{tr}(n,\tau) &=& \frac{A}{\gamma^2}
 \int_0^{\tau} \!\!\!\!
du \, g(u)  \left[ F_n(2\tau-u) -  F_n(u)  \right] 
\nonumber \\
\label{H_F_tr}
\end{eqnarray}
We note that Eq.~\eqref{Hss} and~\eqref{H_F_tr} reduce to the steady state and 
transient MSDs \eqref{app:res_ss} and \eqref{app:res_tr} when $n=0$. This 
follows from
\begin{eqnarray}
    F_0(u) = \int_0^u G(0,t) \, dt = 
    \int_0^u \sqrt{\frac{\tau_0}{4\pi t}} \, dt = \sqrt{\frac{\tau_0 u}{\pi}}
\end{eqnarray}

\subsection{Thermal dynamics}
\label{subsec:equil}
We first verify how the above result for the steady state displacement 
correlation function captures the well-known tension propagation dynamics 
$m(\tau) \sim \tau^{1/2}$ for the equilibrium Rouse model. Steady state 
and transient cases give $H_{ss}(n,\tau) = 2k_BT F_n(\tau)/\gamma$ and 
$H_{tr}(n,\tau) = k_BT F_n(2\tau)/\gamma$ which give essentially the 
same scaling, therefore we restrict to discussing the former only. At short 
times and fixed $n$ the leading behavior for $\tau \ll n^2 \tau_0/2$ is
\begin{eqnarray}
    H^{(th)}(n,\tau) &=& \frac{2 k_B T}{\gamma} F_n(\tau) \sim 
    \frac{\tau^{3/2}}{n^2} \, e^{-\displaystyle{\frac{n^2 \tau_0}{4\tau}}}
    \label{H_th_1}
\end{eqnarray}
where we used $A=2\gamma k_BT$ for thermal noise and used the label ``th'' to
denote the thermal equilibrium dynamics. The short time behavior is obtained 
by the asymptotic expansion of $\text{Erfc}(x)$ for large $x$, which is 
$\text{Erfc}(x) \approx (x \sqrt{\pi})^{-1} e^{-x^2} [1 - 1/(2x^2) + \ldots]$. 
For long times $\tau \gg n^2 \tau_0/2$,  the expansion of Eq.~\eqref{def_Fn} with 
respect to $\alpha(\tau) \ll 1$ leads to 
\begin{eqnarray}
    H^{(th)}(n,\tau) &\approx& \frac{k_BT \tau_0}{\gamma} 
    \left[ \sqrt{\frac{4 \tau}{\pi \tau_0}} - n \right]
    \label{H_th_2}
\end{eqnarray}

\begin{figure}[t!]
    \centering
    \includegraphics[width=1\linewidth]{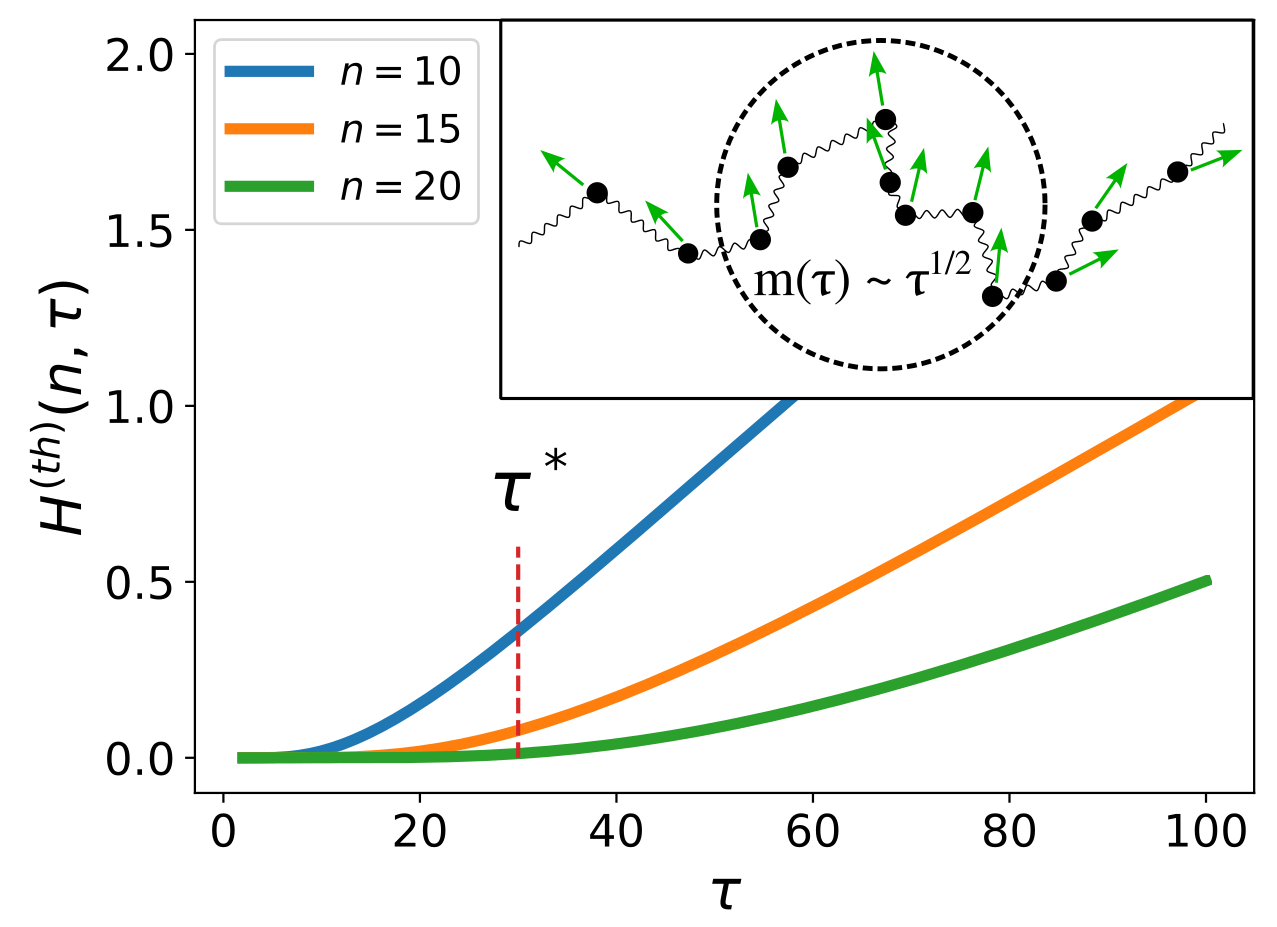}
    \caption{Plot of the short time scale behavior of the displacement correlator
    for thermal equilibrium dynamics $H^{(th)}(n,\tau)$ vs. $\tau$ for three values of $n$. 
    This quantity for small $\tau$ is given by Eq.~\eqref{H_th_1}. At a time scale $\tau^*$
    the displacement of monomers within a separation $n \leq 10$ 
    start showing correlated motion (blue curve)
    while
    those with separation $n \geq 20$ are 
    weakly correlated (green curve).
    Inset: The number of monomers which at
    time $\tau$ perform correlated motion grows
    as $m(\tau) \sim \tau^{1/2}$, according to
    the tension propagation mechanism (see text).
    }
    \label{fig:front_prop}
\end{figure}

Figure~\ref{fig:front_prop} shows a plot of the short time scale behavior 
of $H^{(th)}(n,\tau)$ for three values of $n$. These results show that while 
two distal monomers on the chain are uncorrelated in their motion on short time 
scales, they start to move together on longer time scale~\cite{Katayama_2025}. 
One can estimate the time at which correlated motion sets in from the
condition $H^{(th)}(n,\tau) = const$. This is obtained from the short time 
behavior \eqref{H_th_1} which is dominated by the exponential term leading
to the expected tension propagation dynamics $m(\tau) \sim \tau^{1/2}$.
Note that at very long times \eqref{H_th_2} reproduces the tagged monomer 
MSD scaling as $\sqrt{t}$ with an $n$-independent prefactor. All monomers
within the tension propagation front perform correlated motion and the
effect of their distance is captured by the offset term $-n$ in 
\eqref{H_th_2}. 

\subsection{Active dynamics}
\label{subsec:active}

We now calculate $H(n, \tau)$ for an active polymer kicked by persistent noise 
with characteristic time $\tau_A$; the correlation can be exponential 
$g(u)= e^{-u/\tau_A}$ or more slowly decaying $g(u) = (1 + (u/\tau_A)^{\alpha})^{-1}$ 
with $\alpha < 1$. In particular, we are interested in the short time scale regime 
$\tau \ll \tau_A$, where $g(u) \simeq 1$. We first consider the steady state, then, 
proceed to the transient case. With two independent noise sources given by Eq.~\eqref{f_ss}, 
the displacement correlation function is represented as the sum of two contributions. 
The thermal contribution is given by $H^{(th)}(n,\tau)$ discussed in \eqref{subsec:equil}.
We discuss now separately steady state and transient regimes.

\begin{figure}
    \centering
    \includegraphics[width=0.92\linewidth]{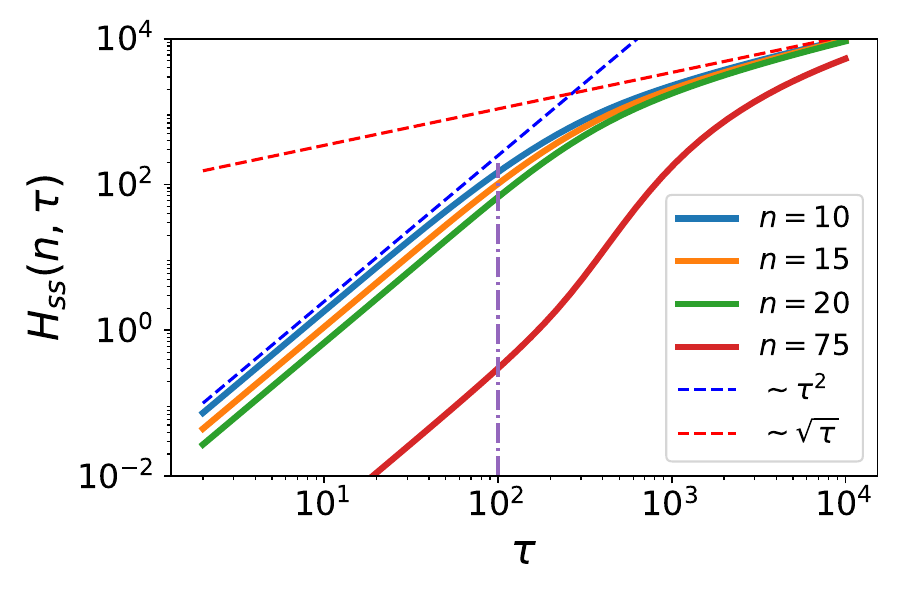}
    \caption{Plot of the steady state displacement correlation $H_{ss}(n,\tau)$
    vs. $\tau$ for a few values of $n$. This quantity is calculated numerically
    from Eq.~\eqref{Hss} using $g(u)=\exp(-u/\tau_A)$. The values used are $\tau_0=1$
    and $\tau_A=100$ (dot-dashed line). The dashed lines show the $\sim \tau^2$ 
    and $\sim \sqrt{\tau}$ behavior expected at short ($\tau \ll \tau_A$) and 
    long ($\tau \gg \tau_A$) time scales.}
    \label{fig:H_active}
\end{figure}

\paragraph{Steady state}
Using Eq.~\eqref{Hss} with $g(u)=1$ for $\tau < \tau_A$ and $g(u)=0$ for 
$\tau > \tau_A$, the active contribution on the short time scale ($\tau \ll \tau_A$) 
is approximated as
\begin{eqnarray}
&&H_{ss}(n,\tau) \simeq \frac{A}{\gamma^2}\int_0^{\tau_A} du \,  
[ F_n(\tau+u) -  2F_n(u)  \nonumber \\ 
&&+ F_n(|\tau-u|)   ] 
= \frac{A}{\gamma^2} [ E_n(\tau_A + \tau) - 2 E_n(\tau_A) 
\nonumber \\
&&+ E_n(\tau_A -\tau)] 
  \simeq  
\frac{A}{\gamma^2} \, G(n,\tau_A) \,  \tau^2 
\nonumber \\
&&= 
\frac{A}{\gamma^2} \sqrt{\frac{\tau_0}{4\pi\tau_A}} \, \,
e^{\displaystyle{-\frac{\tau_0 n^2}{4\tau_A}}} \, \tau^2 
\label{H_F_s_s}
\end{eqnarray}
where we have defined $E_n(u) \equiv \int_0^u dw \, F_n(w)$. Recall that $F_n(t)$ 
is the integral in time of the Gaussian propagator \eqref{def_Fn}, therefore $E_n(t)$ 
is obtained by integrating $G(n,t)$ twice in time. As a consequence the Gaussian 
propagator is the second time derivative of $E_n$
\begin{eqnarray}
    G(n,\tau) = \frac{d^2 E_n(\tau)}{d\tau^2}.
\end{eqnarray}
Expanding $E_n(\tau_A \pm \tau)$ in the second line of \eqref{H_F_s_s} for 
$\tau \ll \tau_A$, one gets that at short times $H_a$ is proportional to $\tau^2$, 
with the $n$-dependence in the Gaussian propagator $G(n,\tau_A)$ as prefactor. 
The condition $G(n,\tau_A) \approx 1$ gives
\begin{eqnarray}
    n \lesssim m_A \equiv \sqrt{\frac{2\tau_A}{\tau_0}}
\end{eqnarray}
which coincides with Eq.~\eqref{m_A} for the standard Rouse model $\eta=2$. 
$m_A$ is the size of steady domain for the cooperative motion. All monomers 
within the range $m_A$ from a given monomer perform a correlated displacement 
with it. 
Importantly, this size $m_A$ is independent of time scale $\tau$ in the 
steady state maintained by the correlated active noise in the time scale 
shorter than the noise decorrelation time. While the final result of \eqref{H_F_s_s} 
was obtained by approximating the exponential AOU noise as a step function 
the results holds in general. Figure~\ref{fig:H_active} shows $H_{ss}(n,\tau)$ 
from numerical estimates of Eq.~\eqref{Hss} with $g(u)=\exp(-u/\tau_A)$.
The short time scale behavior matches indeed the $\sim\tau^2$ prediction of 
Eq.~\eqref{H_F_s_s}. Note that at long time scales $\tau \gg \tau_A$ the 
displacement correlation scales as $H_{ss}(n,\tau) \sim \tau^{1/2}$, which is 
the same as the equilibrium dynamics \eqref{H_th_2}. This is the expected 
behavior as at this time scales the effect of noise persistence (operating up 
to time scales $\approx \tau_A$) can be neglected and the noise can be approximately 
considered as $\delta$-correlated.
We note that recent studies of active polymers have also found collective excitations
generated by temporally patterned noise, that are similar to those discussed 
here \cite{goyc24}.

\paragraph{Transient}
The active noise is switched on at time $s$ at which polymer assumes its 
equilibrium configuration under thermal noise. The noise is thus given 
by Eq.~\eqref{f_tr}. Similarly to the steady state case, the displacement 
correlation function has thermal and active contributions with thermal one 
discussed in Sec.~\ref{subsec:equil}. For the active par we obtain from 
Eq.~\eqref{H_F_tr} at short times $\tau \ll \tau_A$
\begin{eqnarray}
H_{tr}(n,\tau) =  \frac{A}{ \gamma^2} \left[ E_n(2\tau) - 2E_n(\tau) \right]  
\label{H_a_tr}
\end{eqnarray}
where we approximated $g(u) \approx 1$ for $\tau \ll \tau_A$.
In stark contrast to Eq.~\eqref{H_F_s_s} for the active steady state, where the 
spatial correlation is governed by $G(n, \tau_A)$, thus, no dependence on the 
time scale $\tau$, we find $\tau$-dependence in the transient case (recall 
$F_n(\tau), E_n(\tau)$ are obtained from $G(n,\tau)$ by integration).

To summarize, for a given time scale $\tau$, one can define the size $m(\tau) =  
(2\tau/\tau_0)^{1/2}$ such that $H_{th}(n,\tau)$ vanishes for $ n  \gtrsim m(\tau)$. 
The same property, i.e., the vanishing correlation for $ n  \gtrsim m(\tau) $ applies 
to the active contribution in the transient case given in Eq.~\eqref{H_a_tr}. However, 
it does not hold for the active contribution in the steady state, where the time scale 
independent domain with size $m_A(\tau_A)$ is steadily formed. These features are 
correctly captured in the compounding formula through Eq.~\eqref{m_tau_general}.

\section{Summary and Discussions}
\label{sec:conclusion}

Particles with spontaneous activity or kicked by non-thermal noise behave differently 
from passive particles in thermal equilibrium. When such particles are connected into 
a chain, the 
resulting system exhibits a rich variety of dynamics due to the combined 
effect of the activity and the elastic connectivity, which is described by the active 
polymer model at a coarse-grained level. In this paper, we have provided a comprehensive 
discussion on the dynamics of active polymers by introducing and analyzing a compounding 
formula~\eqref{CF2} in the context of a generalized active Rouse model. The formula 
relates the MSD of a tagged polymer locus to that of an individual monomer modulated 
by the effect of the chain connectivity, thereby disentangling the complex dynamical 
behavior into two distinct contributions which are simpler to analyze and understand. 
We have verified that the formula correctly predicts the behavior of the MSD for a 
series of solvable Rouse-like models using different type of active noises.

We emphasize that, although the ideas behind the compounding formula are often 
implicitly invoked to account for the anomalous dynamics of the equilibrium Rouse 
model, to the best of our knowledge no such formula has been explicitly formulated 
or employed in polymer dynamics, beyond thermal systems. Our present formulation 
provides a useful way of thinking to analyze the complex dynamics of active polymers 
kicked by persistence noises. Prior studies \cite{vand15,osma17,Gompper_2020} 
have elucidated the intricate dynamical behavior of active polymeric systems, 
demonstrating the emergence of both subdiffusive and superdiffusive MSD scaling 
regimes. In these works, the MSD was obtained from model-based calculations; 
however, it remained challenging to predict the corresponding scaling exponents 
or to anticipate crossover behavior, since the available analytical expressions 
typically involve infinite summations over Rouse modes. In contrast, the 
compounding-formula framework yields transparent scaling predictions and clearly 
indicates that crossovers between distinct scaling regimes can arise solely from 
the contribution of the isolated-monomer term or from modifications of the 
connectivity factor.

\subsection{Distinct scaling for transient and steady state dynamics in active
systems}

We have pointed out the importance of the protocol, which affects the dynamics. 
Specifically, we have defined the steady state and the transient scenarios depending 
on the history of system preparation and how the measurement is performed.
Although applied to both equilibrium and active polymers (and any other systems with 
restoring force), the statement is particularly true for the latter. Indeed, we have 
shown that while these two protocols share the same MSD scaling for the equilibrium 
polymer, protocol-dependent distinct scalings emerge for the active polymer. 
In this way, one can resolve conflicting observations/predictions in the literature 
on the dynamics of active Rouse polymer $\langle \Delta z(n,\tau)^2 \rangle 
\sim \tau^{2}$~\cite{osma17,eise17,Gompper_2020} or $\langle \Delta z(n,\tau)^2 \rangle 
\sim \tau^{1.5}$~\cite{vand15,vand17,Sakaue_2017,Put_2019}.  

\subsection{Interpretation of ${\cal A}$ and ${\cal B}$ from the real
space analysis}
\label{conc:interpret}

In our real space analysis presented in Sec.~\ref{sec:real_space_analysis}, we 
introduced the displacement decomposition using the terms ${\cal A}_n$ 
and ${\cal B}_n$ as given in Eq.~\eqref{dz-decomposition}. This provides a 
convenient framework for a transparent and systematic discussion of the dynamics 
of active polymers. In particular, it makes a clear distinction between the 
steady state and the transient protocols in the most natural way. We now discuss 
the meaning of each term ${\mathcal A}_n(\tau)$, ${\mathcal B}_n(\tau)$ in the 
decomposition~\eqref{dz-decomposition}.

While Eqs.~\eqref{defA} and \eqref{defB} express these two terms as integrals
in the stochastic noise, it is convenient to rewrite them in a different form.
Assuming the configuration $z(n,s)$ at time $s$ is given, we use the solution~\eqref{gen_sol} to write the displacement of the monomer position $\Delta z_n(s \rightarrow t; s) = z(n,t;s)-z(n,s)$ in the sebsequent time interval $\tau = t-s$
\begin{eqnarray}
\Delta z_n(s \rightarrow t; s) ={\mathcal A}_n(\tau) + \langle z(n,t;s) \rangle_{s \rightarrow t}  - z(n,s)
\end{eqnarray}
where 
\begin{eqnarray}
 \langle z(n,t;s) \rangle_{s \rightarrow t} = \int dn' G(n-n', t-s)z(n',s)  \label{Dz_det}
\end{eqnarray}
We observe that while ${\mathcal A}_n(\tau)$, as defined in \eqref{defA}, represents the stochastic evolution due to the noise, Eq.~\eqref{Dz_det}
describes the deterministic evolution, where the notation $\langle \cdots \rangle_{s \rightarrow t}$ designates taking the average over the noise $f(n',t')$ during the interval $[s, t]$. 
To compare with our definition of the displacement decomposition \eqref{dz-decomposition},  we rewrite $z(n,s)$ as 
\begin{eqnarray}
 &&z(n,s;-T_\infty)  \nonumber \\
 &&= \frac{1}{\gamma}\int dn' \int_{-T_{\infty}}^s dt' \, G(n-n', s-t') f(n',t') 
 \label{z_s_t_0}
\end{eqnarray}
 to explicitly represent that the configuration at time $s$ is obtained through the past time evolution from the initial configuration $z(n,-T_{\infty})$, where we drop the initial configuration dependence by assuming $T_\infty$ longer than any time scale of 
the system.
We then find
\begin{eqnarray}
{\mathcal B}_n(\tau) =  \langle z(n,t;-T_\infty) \rangle_{s \rightarrow t}
- \ z(n,s;-T_\infty) 
\label{def_B_2}
\end{eqnarray}
where
\begin{eqnarray}
&&\langle z(n,t;-T_\infty) \rangle_{s \rightarrow t} \nonumber \\
&&= \int dn'  \, G(n-n', t-s) z(n',s;-T_\infty) \label{z_bar}
\end{eqnarray}

\begin{figure}[t!]
    \centering
    \includegraphics[width=1\linewidth]{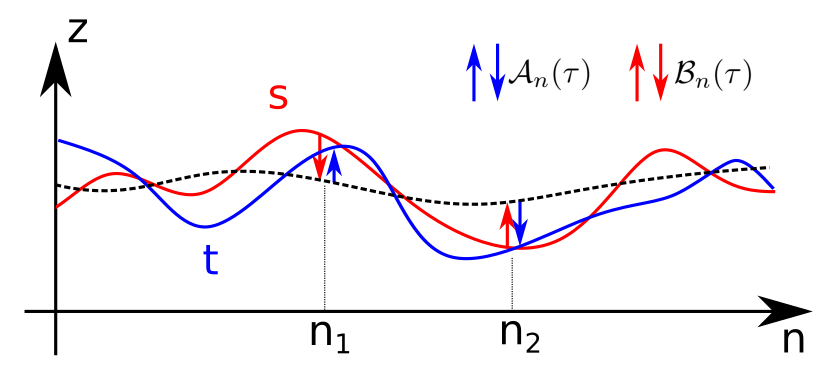}
    \caption{Disentanglement of the contributions ${\cal A}_n$ 
    and ${\cal B}_n$ to the total displacement $\Delta z_n(s 
    \rightarrow t; -T_\infty) = z(n,t;-T_\infty)-z(n,s;-T_\infty)$. 
    In the graph $z(n,s;-T_\infty)$ and $z(n,t;-T_\infty)$, plotted 
    as a function of the continuous monomer index $n$, are shown as 
    red and blue curves, respectively. The dotted line is obtained 
    from the red curve via a convolution with the Gaussian propagator 
    $G(n,t)$, see~\eqref{def_B_2},~\eqref{z_bar}. ${\cal B}_n$ (red arrow) connects the $z(n,s;-T_\infty)$
    to the dotted curve, which is a smoother curve as expected for a 
    diffusive process. ${\cal A}_n$ (blue arrow) is the contribution to the displacement
    connecting the dotted curve to $z(n,t;-T_\infty)$. Arrows pointing up 
    or down indicate positive or negative contributions to the displacement. 
    Since the profile $z(n,s;-T_\infty)$ (red curve) is created by the noise 
    sequence up to time $s$, the physical nature of the subsequent relaxation 
    and the noise persistence indicates that blue and red arrows tend to be 
    oppositely oriented.}
    \label{fig:AnBn}
\end{figure}

Equation~\eqref{def_B_2} provides an alternative representation to the expression given
in \eqref{defB}. 
As $\langle z(n,t;-T_\infty) \rangle_{s \rightarrow t}$ in Eq.~\eqref{z_bar}
solves the Rouse equation \eqref{eq:Rouse} in the deterministic limit
$f=0$, the term ${\mathcal B}_n(\tau)$ describes a relaxation dynamics of 
diffusive type from a given configuration $z(n,s;-T_\infty)$ at time $s$.
As Eq.~\eqref{z_s_t_0} shows, however, that $z(n,s;-T_\infty)$ is a stochastic variable 
depending on the noise $f(n',t')$ acting during the interval $[-T_\infty,s]$.
We know, on the other hand, the term ${\mathcal A}_n(\tau)$ represents the stochastic displacement caused by the
noise during the interval $[s,t]$, see Eq.~\eqref{defA}. Therefore, both ${\mathcal A}_n(\tau)$ and ${\mathcal B}_n(\tau)$ are
functionals of the noise $f(n',t')$ but from different time intervals.
These features are best
shown in Fig.~\ref{fig:AnBn} which illustrates the evolution
from a conformation $z(n,s;-T_\infty)$ towards $z(n,t;-T_\infty)$, shown as red
and blue curves respectively. The displacement can be split into two parts
as indicated by the colored arrows, where the dotted path is obtained from
the red one via the convolution with the Gaussian propagator $G(n,t)$, 
see~\eqref{def_B_2},~\eqref{z_bar}. We refer to this part as relaxation dynamics and it is
analogous to a diffusion process smoothening the initial profile 
$z(n,s;-T_\infty)$. 
In the absence of noise in the time interval $[s,t]$, one would have ${\cal A}_n(\tau)=0$, see \eqref{defA}, and there would be only relaxation dynamics.

\subsection{Cross correlations $\langle {\cal AB}\rangle$}

In the transient scenario, the active noise is absent up to time $s$. Therefore, 
there is no active contribution to ${\mathcal B}_n(\tau)$. The remaining term 
${\mathcal A}_n(\tau)$ thus represents the genuine transient contribution. In 
the steady state scenario, the existence of the ${\mathcal B}_n(\tau)$ term adds 
in the MSD not only $\langle {\mathcal B}_n^2(\tau) \rangle$ term but also the 
cross term $2 \langle  {\mathcal A}_n(\tau) {\mathcal B}_n(\tau) \rangle$. The 
latter represents the fact that the relaxation dynamics of the system couples 
to the noise induced stochastic evolution. This is because the relaxation starts 
from 
$z(n,s;-T_\infty)$, which is
created by the past noise history ($<s$).
This is correlated to the noise in the interval $[s,t]$, since the
noise is colored.
We note that it can be shown that $\langle  {\mathcal A}_n(\tau) 
{\mathcal B}_n(\tau) \rangle <0$, see comment below Eq.~\eqref{int_dom}.
The physical explanation is that ${\mathcal A}$ and ${\mathcal B}$ terms tend 
to have opposite signs for the following reason: the persistence of the noise tends 
to maintain a profile at time $t$ similar to that at time $s$, opposing the 
relaxation dynamics, see blue and red arrows in Fig.~\ref{fig:AnBn}.
For a polymer in thermal equilibrium, no such correlation exists due to the white noise nature of the thermal noise, see further discussion on the thermal dynamics in Appendix~\ref{sec:equil-final:real-space}.

Our calculation in Sec.~\ref{sec:real_space_analysis} has shown that the sum of 
the past noise contributions $\langle {\mathcal B}_n^2(\tau) \rangle + 
2 \langle  {\mathcal A}_n(\tau) {\mathcal B}_n(\tau) \rangle$ cancels the 
transient $\langle {\mathcal A}_n^2(\tau) \rangle$ term, changing the scaling 
from $\langle \Delta z^2(n, \tau)\rangle_{tr} \sim \tau^{2 - 1/\eta}$ to 
$\langle \Delta z^2(n, \tau)\rangle_{ss}\sim \tau^2$. 
In the compounding formula, such a feature is encoded in the connectivity 
factor~\eqref{m_tau_general}. As our calculation of the displacement correlation 
function shows, if the noise has some persistence, such a past noise history is 
responsible for the formation of the steady domain with size $m_A$.

\subsection{Fast/slow dynamics}
\label{subsec:Conslucion_fast_slow_dynamics}
Finally, we emphasize the transient dynamics is faster than the steady state 
dynamics $\langle \Delta z^2(n, \tau)\rangle_{tr} > \langle \Delta z^2(n, \tau)\rangle_{ss}$ 
for the active polymer despite the larger exponent for the latter, see Fig~\ref{fig:compare}. 
The result might be surprising if one compares it with the equilibrium polymer 
for which the opposite is true, i.e., $\langle \Delta z^2(n, \tau)\rangle_{tr} < 
\langle \Delta z^2(n, \tau)\rangle_{ss}$. The latter inequality for the equilibrium 
polymer is most readily understood through our discussion in Appendix~\ref{sec:equil-final}, 
where we present a compelling argument based on the ${\mathcal A}_n(\tau) \ {\mathcal B}_n(\tau)$ 
decomposition.
The inequality for the active polymer is also readily understood by the physical picture 
offered by the compounding formula. Indeed, in the transient scenario, each monomer 
can move independently up to the time scale $\tau_0$ at which $m(\tau_0) = 1$. From then 
on, $m(\tau)$ grows according to Eq.\eqref{m_tau_general} (top), leading to the connectivity 
induced gradual slowing down, which is described by the transient MSD exponent. In contrast, 
the steady domain size $m_A$ ~\eqref{m_A} is preformed in the steady state scenario, so 
that each monomer moves collectively with neighboring $m_A$ monomers from the beginning. 
In other words, the tagged monomer dynamics is understood as that of the center-of-mass 
mode of a subchain made from $m_A$ monomers. This naturally explains the ballistic MSD 
exponent irrespective of the polymer connectivity $\eta$, which persists up to the time 
scale $\tau_A$. At $\tau \simeq \tau_A$ when the persistence of the noise is lost, the 
two MSD $\langle \Delta z^2(n, \tau)\rangle_{tr}$, $ \langle \Delta z^2(n, \tau)\rangle_{ss}$ 
align, and the subsequent dynamics becomes effectively the thermal one with the MSD exponent 
$(\eta-1)/\eta$, see Eqs.~\eqref{MSD_CF_prediction_tr} and \eqref{MSD_CF_prediction_ss} albeit 
with a higher effective temperature or a larger diffusivity. Thus, on such long time scales, 
our discussion in Appendix~\ref{sec:equil-final:real-space} applies, which indicates the relation 
$\langle \Delta z^2(n, \tau)\rangle_{tr} < \langle \Delta z^2(n, \tau)\rangle_{ss}$. 
The transition from the early $\langle \Delta z^2(n, \tau)\rangle_{tr} > 
\langle \Delta z^2(n, \tau)\rangle_{ss}$ to the late $\langle \Delta z^2(n, \tau)\rangle_{tr} 
< \langle \Delta z^2(n, \tau)\rangle_{ss}$ is indeed seen in Figure~\ref{fig:compare}.
Finally, if the noise has a long 
persistence as in Eq.~\eqref{g_t_pow}, it will leave its trail in the long time scale MSD 
scaling as in Eq.~\eqref{MSD_CF_prediction_pw_noise}.

\begin{acknowledgments}
T.S thanks J. Prost and G.V. Shivashankar for useful discussions. 
This work is supported by JSPS KAKENHI (Grant No. JP23H00369 and JP24K00602).
\end{acknowledgments}

\appendix

\section{Equilibrium dynamics: transient vs. steady-state}
\label{sec:equil-final}
In the main text, we assumed the form $f(n,t) = f_{th}(n,t) + f_a(n,t)$ 
(steady state scenario) or $f(n,t) = f_{th}(n,t) + \theta(t-s)f_a(n,t)$ 
(transient scenario) for stochastic noise, and focused on the contribution 
to the dynamics of the active component $f_a(n, t)$. 
Here we present the calculation for the contribution of the thermal component 
$f_{th}(n,t)$ both in the normal mode analysis in Sec.~\ref{sec:normal_mode_analysis} 
and the real space approach in Sec.~\ref{sec:real_space_analysis}, where we set 
$A = 2 \gamma k_BT$ and $g(t) = \delta (t)$ for the noise auto-correlation. 
While the normal mode analysis is a standard textbook content~\cite{doi88}, 
the real space approach provides us with a concise view on the relation between 
steady state and transient MSDs.

The transient process generally depends on the initial state. In the same manner 
as the transient active noise, we define the thermal noise sequence in the transient 
scenario as $ \theta(t-s)f_{th}(n,t)$. This amounts to prepare the system in the 
ground state configuration at time $s$, or equivalently, to adopt the ``flat" 
initial configuration.
In the Rouse polymer, this corresponds to a collapsed conformation. In more 
realistic polymer with self-avoidance, one can take the stretched conformation in 
a direction perpendicular to $z$ axis. 
 
It may be useful to give another example to illustrate such a transient process. 
Let us consider the interface that grows by random deposition of particles. Upon falling 
on the top of the interface, deposited particles are allowed to relax to the lowest 
height neighboring site. In the continuum limit, this process is described by 
Eq.~\eqref{eq:Rouse}, in this context, called Edwards-Wilkinson equation, 
where $z(n, t)$ represents the height of the interface on the one-dimensional 
substrate position $n$ at time $t$ and $k$ plays the role of surface tension. 
The noise is usually assumed to be uncorrelated white noise. The model is thus 
equivalent to the equilibrium Rouse dynamics. In the transient scenario, the 
system starts from a flat interface $z(n,0)=0$ for all $n$, and we monitor the 
height increment $\Delta z_n(0 \rightarrow t; 0) \equiv z(n,t;0) - z(n,0;0)$, 
see Eq.~\eqref{Dz1}, where we set $s=T_\infty=0$, $\tau=t$.

\subsection{Normal mode analysis}
 We first obtain from Eq.~\eqref{C_p_t0}
\begin{eqnarray}
&&C_p(t,s;t_0) = \langle Z_p(t_0)^2 \rangle 
e^{-(t+s-2t_0)/\tau_p}  \nonumber \\
&& \quad + \frac{k_BT \tau_p}{2 N \gamma}\left( e^{-|t-s|/\tau_p} 
- e^{-(t+s-2t_0)/\tau_p} \right) 
\label{C_p_th} 
\end{eqnarray}
\paragraph{Steady state}
Letting $t_0 \rightarrow - \infty$, we obtain the steady state, i.e., equilibrium 
in the present case, correlation function
\begin{eqnarray}
C^{(ss)}_p(\tau) = 
\frac{k_BT \tau_p}{2 N \gamma}
e^{-\tau/\tau_p}  
\label{C_ss}
\end{eqnarray}
The same result is obtained by assuming that the system is thermalized already 
at $t=t_0$ and using
the equipartition theorem $\langle Z_p(t_0)^2\rangle = k_BT/(2 N k_p) =
k_BT \tau_p/(2N\gamma)$.
Using Eqs.~(\ref{MSD_mode_ss}) and~(\ref{C_ss}), we calculate the steady 
state MSD of a tagged monomer as
\begin{eqnarray}
\langle \Delta z^2(n,\tau) \rangle_{ss} 
=  \sum_{p =1}^N  \frac{2  k_BT \tau_p}{N \gamma} \left( 1 - e^{ -\tau /\tau_p }\right) 
\label{MSD_ss_eq}
\end{eqnarray}
The scaling structure can be evaluated as follows.
From the relation $\tau/\tau_p = \tau p^{\eta}/\tau_R \simeq 1$, we define the time-scale 
dependent characteristic mode number $p^*(\tau) \equiv (\tau_R/\tau)^{1/\eta}$, which 
acts as the lower cut-off of the summation. We then approximate the summation in 
Eq.~\eqref{MSD_ss_eq} by integral, where the range of integral is rationalized by noting 
the time-scale of our interest $\tau_0 \ll \tau \ll \tau_R  \Leftrightarrow 1 \ll p^* \ll N$. 
Using $\tau_p = \tau_R/p^\eta = \tau_0 (N/\pi p)^\eta$ we find
\begin{eqnarray}
\langle \Delta z^2(n,\tau) \rangle_{ss} &=& \!\sum_{p =1}^N  
\frac{2  k_BT \tau_0 N^{\eta-1}}{\gamma (\pi p)^{\eta}} \!
\left[ 1 - \exp{\left( - \frac{\tau p^{\eta}}{\tau_R} \right)}\right] 
\nonumber \\
&\simeq&  \frac{2  k_BT \tau_0 N^{\eta-1}}{\gamma \pi^{\eta}} \int_{p^*}^{\infty} dp \   p^{-\eta}  \nonumber \\
&\simeq&\frac{2k_BT \tau_0}{(\eta-1) \pi \gamma} \left( \frac{\tau}{\tau_0}\right)^{(\eta-1)/\eta}
\label{tag_mon_th_2}
\end{eqnarray}
For $\eta=2$, we obtain the classical Rouse scaling $\langle \Delta z^2(n,\tau) \rangle_{ss} \sim \tau^{1/2}$.

\paragraph{Transient}
The transient MSD is calculated from Eqs.~\eqref{MSD_mode_tr} and~\eqref{C_p_th};
\begin{eqnarray}
\langle \Delta z^2(n,\tau) \rangle_{tr} 
&=& 2 \sum_{p=1}^N \left[ \langle Z^2_p(0) \rangle (1- e^{-\tau/\tau_p})^2 \right.
\nonumber \\
&&
+ \left. \frac{k_BT \tau_p}{2 N \gamma} ( 1-e^{-2 \tau/\tau_p})\right]  \nonumber \\
\label{MSD_tr_eq_0}
\end{eqnarray}
If we assume the system is prepared in the thermal equilibrium state, applying 
the equipartition theorem for the initial condition, the above expression reduces 
to the steady-state MSD given by Eq~\eqref{MSD_ss_eq}. For different initial 
conditions away from equilibrium, the transient relaxation process toward 
equilibrium state set in. For a ``flat" initial 
state $\langle Z_p(0)^2\rangle =0$ for all $p$, we obtain
\begin{eqnarray}
\langle \Delta z^2(n,\tau) \rangle_{tr} &= & 
\sum_{p=1}^N \frac{k_BT \tau_p}{ N \gamma} ( 1-e^{-2 \tau/\tau_p})
\label{MSD_tr_eq}
\end{eqnarray}
The transient MSD~\eqref{MSD_tr_eq} can be evaluated in the same manner as 
the steady state MSD. Indeed, as the expressions~\eqref{MSD_ss_eq} 
and~\eqref{MSD_tr_eq} are similar, we expect the same scaling structure 
$\langle \Delta z^2(n,\tau) \rangle_{tr} \sim \tau^{(\eta -1)/\eta}$ with 
a slight difference in their prefactors. 
We will see the exact relation between these MSDs below for the Rouse model $\eta=2$.

\subsection{Real space analysis}
\label{sec:equil-final:real-space}
We recall our definition of the displacement, see Eq.~\eqref{Dz1};
\begin{eqnarray}
    \Delta z_n(s \rightarrow t; t_0) \equiv z(n,t;t_0) - z(n,s;t_0)
\end{eqnarray}
\paragraph{Steady state}
In the steady state scenario in equilibrium dynamics, the system is aged 
($t_0 = -T_\infty$ with $T_\infty \rightarrow \infty$) and settled in its steady state in the presence 
of the uncorrelated white noise, and we monitor the displacement during the 
time interval $\tau = t-s$, i.e., $\Delta z_n(s \rightarrow t; -T_\infty) 
\equiv z(n,t;-T_\infty) - z(n,s;-T_\infty)$.
Using Eqs.~\eqref{MSD_A-B}, ~\eqref{resA2}, ~\eqref{resB2} and noting the absence of cross term  $\langle {\cal A}_n {\cal B}_n \rangle =0$ due to the 
white noise nature of the thermal noise, we obtain the steady state MSD
\begin{eqnarray}
\langle \Delta z(n,\tau) ^2 \rangle_{ss} 
&=& \langle  {\cal A}_n(\tau)^2\rangle + 
\langle  {\cal B}_n(\tau)^2\rangle 
 = \frac{2k_BT}{ \gamma} \sqrt{\frac{\tau_0 \tau}{\pi}} 
\nonumber \\
\label{MSD_eq_ss}
\end{eqnarray} 
This is the ordinary MSD observed for Rouse model in thermal equilibrium.
In this case, both ${\cal A}_n(\tau)$ 
and ${\cal B}_n(\tau)$ are nonzero with the latter carrying the information 
of past history of the noise, see Eq.~\eqref{defB}.
Although there is no correlation between ${\cal A}_n(\tau)$ and ${\cal B}_n(\tau)$, they represent the stochastic evolution (fluctuation) and the relaxation, respectively, thus, are connected as a consequence of the fluctuation-dissipation theorem. Indeed, they behave in a similar way with $\langle {\cal A}_n(\tau)^2 \rangle \sim \langle {\cal B}_n(\tau)^2 \rangle \sim \tau^{0.5}$.
 

\paragraph{Transient}

For a ``flat" initial configuration $z(n,t_0)=0$ for all $n$ at $t_0=0$, 
we obtain from the solution~\eqref{gen_sol} of Rouse equation of motion, 
\begin{eqnarray}
&&\Delta z_n(0 \rightarrow t;0) = \nonumber \\
&& \frac{1}{\gamma}  \int dn' \int_0^t dt' \ G(n-n',t-t') \ f(n',t')
\end{eqnarray}
Comparing the above expression with Eqs.~\eqref{dz-decomposition}, \eqref{defA},
\eqref{defB}, we find that the ``flat" initial condition amounts to set 
${\cal B}_n(\tau)=0$.
Therefore, from Eq.~\eqref{MSD_A-B}, the MSD is given by
\begin{eqnarray}
\langle \Delta z(n,\tau) ^2 \rangle_{tr} &=& \langle  {\cal A}_n(\tau)^2\rangle 
=
\frac{k_BT}{ \gamma} \sqrt{\frac{2\tau_0\tau}{\pi}} 
\label{MSD_eq_tr}
\end{eqnarray} 
Comparing Eqs~\eqref{MSD_eq_ss} and \eqref{MSD_eq_tr}, we conclude 
$\langle \Delta z(n,\tau) ^2 \rangle_{tr} < \langle \Delta z(n,\tau) ^2 \rangle_{ss}$ 
due to the lack of ${\cal B}_n(\tau)$ factor in the transient case, 
while these two cases share the same scaling structure. 
We expect that such a relation between the steady state MSD and the 
transient MSD would be quite common for systems kicked by white noise, 
thus valid for a general polymer model with $\eta \neq 2$.
For a Rouse polymer $\eta=2$, the latter is quantitatively smaller 
by a factor of $\sqrt{2}$ as revealed by the above calculation.


Finally, we point out that the above relation between steady state and transient 
mobilities in equilibrium dynamics is completely altered in the active dynamics. 
As discussed in Sec.~\ref{subsec:Conslucion_fast_slow_dynamics}, 
the active systems on time scale $\tau < \tau_A$ exhibit (i) different MSD scalings in steady state and transient processes, (ii) their magnitude relation is $\langle \Delta z_n(\tau) ^2 \rangle_{tr} > \langle \Delta z_n(\tau) ^2 \rangle_{ss}$ 
that is opposite to that in the equilibrium dynamics. Our physical picture based 
on a compounding formula suggests that these properties are expected to be generic 
on time scale shorter than the active correlation time $\tau_A$.

\section{Details of the calculations of $\langle {\cal A}_n^2\rangle$, 
        $\langle {\cal B}^2\rangle$ and $\langle {\cal A}_n {\cal B}_n\rangle$}
\label{subsec:details}

To derive the final expressions we used \eqref{auto_corr_noise} and
\begin{equation}
    \int dn' \ G(n-n', \tau) \ G(n'-n'', \tau') = G(n-n'',\tau+\tau')
    \label{CGeq}
\end{equation}
which follows from elementary properties of the Gaussian propagator \eqref{def_prop}.

\subsection{Calculation of $\langle {\cal A}^2_n\rangle$}
From the definition of ${\cal A}_n(\tau)$, see Eq.~\eqref{defA}, we have
\begin{widetext}
\begin{eqnarray}
    \langle {\cal A}_n(\tau)^2 \rangle &=& \frac{1}{\gamma^2} \int dn' dn'' \int_s^t dt' dt''
    \ G(n-n',t-t') \ G(n-n'',t-t'') \ \langle f(n',t') f(n'',t'') \rangle
    \nonumber \\
    &=& 
    \frac{A}{\gamma^2} \int_s^t dt'dt'' g(|t'-t''|) \int dn' G(n-n',t-t') \ G(n-n',t-t'')
    \nonumber \\
    &=& 
    \frac{2A}{\gamma^2} \int_s^t dt' \int_s^{t'} dt'' g(t'-t'') \ G(0, 2t - t' -t'')
\end{eqnarray}
\end{widetext}
where we used \eqref{CGeq}.
With the following change of variables $u \equiv t'-t''$ and $w = 2(t-t')$ we can
rewrite the integration domain as
\begin{eqnarray}
    \langle {\cal A}_n(\tau)^2 \rangle &=& \frac{A}{\gamma^2} \int_0^{\tau} du \ g(u) 
    \int_{0}^{2(\tau-u)} \!\!\!\!\! dw \ G(0,u+w)  \label{A_square_G} 
    \nonumber \\
    &=&
    \frac{A}{\gamma^2} \sqrt{\frac{\tau_0}{4\pi}} \int_0^{\tau} du \ g(u) 
    \int_{0}^{2(\tau-u)} \!\!\!\!\!\!\!\!\frac{dw}{\sqrt{u+w}}
    \nonumber \\
    &=& 
    \frac{A}{\gamma^2} \sqrt{\frac{\tau_0}{\pi}} \int_0^{\tau} du \ g(u) 
    \left( \sqrt{2\tau - u} - \sqrt{u} \right) \nonumber \\
\end{eqnarray}
with $\tau \equiv t-s$, which is Eq.~\eqref{resA2}.

\subsection{Calculation of $\langle {\cal B}^2_n\rangle$}
We use the similar variable transformation as for the calculation of $\langle {\cal A}_n^2\rangle$, 
where we set the argument in each $G$ term to be $u+w$; 
\begin{widetext}
\begin{eqnarray}
  \langle {\cal B}_n(\tau)^2 \rangle &=&  
   \frac{A}{\gamma^2} \int_0^{s+T_\infty} \!\!\!\!\!\!\!\! 
        du \ g(u)  
    \left[ \int_{2 \tau}^{2(\tau+(T_\infty+s)-u)}dw  - 2  \int_{
\tau}^{\tau+2(T_\infty+s-u)}dw +  \int_{0}^{2(T_\infty+s-u)} dw \right] G(0,u+\omega) \nonumber
\end{eqnarray}
\end{widetext}
where we employed a short-handed notation for the integral: 
$[\int_a^b dw + \int_c^d dw] G(w) \equiv \int_a^b dw \ G(w) + \int_c^d dw \ G(w)$.
The integration range for $w$ in the above equation can be arranged as
\begin{widetext}
\begin{eqnarray}
&&\left[ \int_{2 \tau}^{2(\tau+(T_\infty+s)-u)}dw  - 2  \int_{
\tau}^{\tau+2(T_\infty+s-u)}dw +  \int_{0}^{2(T_\infty+s-u)} dw \right] G(0,u+w)\nonumber \\
&=&\left[ \int_{0}^{\tau}dw  -   \int_{ \tau}^{2\tau}dw  \right] G(0,u+w)
- \left[ \int_{2(T_\infty +s -u)}^{2(T_\infty +s -u)+\tau}dw  -   \int_{ 2(T_\infty +s -u)+\tau}^{2(T_\infty +s -u+\tau)}dw  \right] G(0,u+w) \nonumber \\
&\stackrel{T_\infty \to +\infty}{=}&\left[ \int_{0}^{\tau}dw  -   \int_{ \tau}^{2\tau}dw  \right] G(0,u+w)
\label{B_square_G}
\end{eqnarray}
\end{widetext}
where in the last line we took the limit $T_\infty \to +\infty$, which leads to the
vanishing contribution from the last term. Substituting the form of $G(0,u)$, we find
\begin{widetext}
    \begin{eqnarray}
\langle {\cal B}_n(\tau)^2 \rangle &\stackrel{T_\infty \to +\infty}{=}&
     \frac{A}{\gamma^2} \sqrt{\frac{\tau_0}{4 \pi}}\int_0^{+\infty} \!\!\!\!\!\!\!\! 
        du \ g(u)  
   \left[ \int_{0}^{\tau}dw  -   \int_{ \tau}^{2\tau}dw  \right] \frac{1}{\sqrt{u+w}}
     \nonumber\\
    &=& 
    \frac{A}{\gamma^2} \ \sqrt{\frac{\tau_0}{\pi}} 
    \int_0^{+\infty} \!\!\!\!\!\! du \ g(u)
    \left( 2 \sqrt{\tau + u} - \sqrt{2\tau + u} - \sqrt{u}\right)
    \end{eqnarray}
\end{widetext}
which is the result given in \eqref{resB2}. 

\begin{figure}[t]
    \centering
    \includegraphics[width=0.98\linewidth]{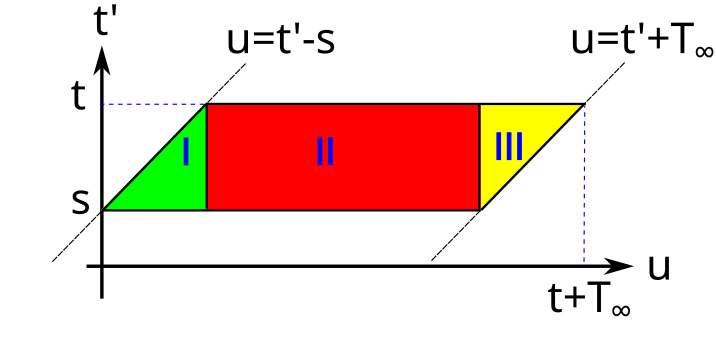}
    \caption{Integration domain in the calculation of $\langle {\cal AB} \rangle$ 
    as function of the variables $u \equiv t'-t''$ and $t'$.}
    \label{fig:AB_domain}
\end{figure}

\subsection{Calculation of $\langle {\cal A}_n {\cal B}_n\rangle$}
From the definition of ${\cal A}_n(\tau), {\cal B}_n(\tau)$, see 
Eqs.~\eqref{defA},~\eqref{defB}, we have
\begin{widetext}
\begin{eqnarray}
    2 \langle {\cal A}_n(\tau) {\cal B}_n(\tau) \rangle &=& \frac{2}{\gamma^2} \int dn' dn'' \int_s^{t} dt' 
    \int_{-T_\infty}^s \!\!\!\!\! dt'' G(n-n',t-t')  \left[ G(n-n'',t-t'') - G(n-n'',s-t'') \right] 
    \nonumber \\
    && \langle f(n',t') f(n'',t'') \rangle 
    =
    \frac{2A}{\gamma^2} \int_s^t dt' \int_{-T_\infty}^{s} dt'' \ g(t'-t'') 
    \left[ G(0,2t-t'-t'') - G(0,s+t-t'-t'') \right] 
    \nonumber\\
    \label{int_dom}
\end{eqnarray}
\end{widetext}
We note that the previous correlator is always negative at all times, which 
follows from the Gaussian propagator inequality $G(0,2t-t'-t'') \leq G(0,s+t-t'-t'')$, 
being $s < t$.

To proceed further we perform a change of variables defining $u \equiv t'-t''$ and
using $u$ and $t'$ as integration variables. The integration domain is then
$s \leq t' \leq t$ and $t'-s \leq u \leq t'+ T_\infty$, which is shown in 
Fig.~\ref{fig:AB_domain}. We split this domain in three parts I, II and III, see
Fig.~\ref{fig:AB_domain}, and proceed with integrating over each domain.
Again, setting the argument in the $G$ term to be $u+w$, we continue the calculation to find
\begin{widetext}
\begin{eqnarray}
     2 \langle {\cal A}_n(\tau) {\cal B}_n(\tau) \rangle_I &=& \frac{2A}{\gamma^2} \int_0^{t-s} du \ g(u) \int_{s}^{u+s} dt' 
     \ \left[ G(0,2t+u-2t') - G(0,s+t+u-2t') \right] 
      \nonumber \\
     &=&  
     \frac{A}{\gamma^2}
     \int_0^{\tau} du \ g(u) \left[ \int_{2(\tau-u)}^{2\tau} dw - \int_{\tau-2u}^{\tau} dw
     \    \right] G(0,u+w) \label{AB_I_G} \\
     &=& 
     \frac{A}{\gamma^2}  \sqrt{\frac{\tau_0}{\pi}} \int_0^{\tau} du \ g(u)
     \left( 
     \sqrt{2\tau + u} - \sqrt{2\tau - u} - \sqrt{\tau + u} + \sqrt{\tau - u}
     \right) 
     \label{AB_I} 
\end{eqnarray}
\begin{eqnarray}
      2 \langle {\cal A}_n(\tau) {\cal B}_n(\tau) \rangle_{II} &=& \frac{2A}{\gamma^2} \int_{t-s}^{s+T_\infty} du \ g(u) \int_s^t dt' 
     \ \left[ G(0,2t+u-2t') - G(0,s+t+u-2t') \right] 
     \nonumber \\
      &=& 
     \frac{A}{\gamma^2}      
      \int_{\tau}^{s+T_\infty} du \ g(u)  
     \ \left[\int_0^{2 \tau} dw - \int_{-\tau}^{\tau} dw   \right] G(0,u+w) \label{AB_II_G} \\
    &\stackrel{T_\infty \to +\infty}{=}& 
     \frac{A}{\gamma^2}  \sqrt{\frac{\tau_0}{\pi}} \int_{\tau}^{+\infty} \!\!\!\!\!\! du \ g(u)
     \left( \sqrt{2\tau + u} - \sqrt{u} - \sqrt{\tau + u} + \sqrt{u-\tau}
     \right)
     \label{AB_II}
\end{eqnarray}
\end{widetext}
Finally $\langle {\cal A}_n(\tau) {\cal B}_n(\tau) \rangle_{III}$ vanishes in the limit $T_\infty \to \infty$.
Summing up \eqref{AB_I} and \eqref{AB_II} one 
gets \eqref{resAB}.

\section{Displacement correlations $H(n,\tau)$}
\label{app:displacement}

The correlation in displacement of two monomers $n$ apart along the chain during 
the time scale $\tau$ can be quantified by the following correlation function;
\begin{equation}
H(n, s \rightarrow t; t_0) \equiv \langle \Delta z_{n_1}(s \rightarrow t; t_0) 
\Delta z_{n_2}(s \rightarrow t; t_0) \rangle
\end{equation}
where $n= |n_1 - n_2|$. 
We set $t_0 = - T_\infty$ with $T_\infty \rightarrow \infty$ and write the displacement 
correlation function in steady state as $H_{ss}(n, \tau)$; the translational invariance along the chain, valid 
for a long chain, and the time-translational invariance in steady state implies that the 
correlation depends only on the separation $n$ and $\tau = t-s$.
We note that by setting $n=0$, the above displacement correlation function reduces to 
the tagged monomer MSD.
Using the decomposition~\eqref{dz-decomposition}, the displacement correlation 
function can be written as
\begin{eqnarray}
H_{ss}(n,\tau) &=& \langle {\cal A}_n(\tau) {\cal A}_0(\tau) \rangle +
2\langle {\cal A}_n(\tau) {\cal B}_0(\tau) \rangle 
\nonumber \\
&+&  \langle {\cal B}_n(\tau) {\cal B}_0(\tau) \rangle
\end{eqnarray}
where we have used $  \langle {\cal A}_n(\tau) {\cal B}_0(\tau) \rangle =  
\langle {\cal B}_n(\tau) {\cal A}_0(\tau) \rangle$, a relation that follows 
from the inversion symmetry as well as the translational invariance along the chain.
To calculate these quantities, we can follow essentially the same step 
detailed above in the calculation of the MSD. 
\begin{widetext}
\begin{eqnarray}
\langle {\cal A}_n(\tau) {\cal A}_0(\tau) \rangle &=&
 \frac{A}{\gamma^2} \int_0^{\tau} du \ g(u) 
    \int_{0}^{2(\tau-u)} dw \ G(n,u+w)  \label{AnA0}\\
  \langle {\cal B}_n(\tau) {\cal B}_0(\tau) \rangle &=&  
  \frac{A}{\gamma^2} \int_{0}^{+\infty} du \ g(u)  \left[ \int_{0}^{\tau}dw  -   
  \int_{ \tau}^{2\tau}dw  \right] G(n,u+w) \label{BnB0} \\
  2\langle {\cal A}_n(\tau) {\cal B}_0(\tau) \rangle &=& 
   \frac{A}{\gamma^2} \int_0^{\tau} du \ g(u) \left[ \int_{2(\tau-u)}^{2\tau} dw - \int_{\tau-2u}^{\tau} dw
     \    \right] G(n,u+w) \nonumber \\
     &+&  \frac{A}{\gamma^2} \int_{\tau}^{+\infty} du \ g(u)  
     \ \left[\int_0^{2 \tau} dw - \int_{-\tau}^{\tau} dw   \right] G(n,u+w) \label{AnB0}
\end{eqnarray}

Breaking the integration range of $u$ in $ \langle {\cal B}_n(\tau) {\cal B}_0(\tau) 
\rangle$ as $\int_0^{+\infty} du = \int_0^{\tau} du + \int_{\tau}^{+\infty}du $, 
we can write and arrange $H(n,\tau)$ as
\begin{eqnarray}
H_{ss}(n,\tau) &=& \frac{A}{\gamma^2}\int_0^{\tau} du \, g(u)  \left[ \int_0^{2(\tau-u)} dw +
\int_{2(\tau-u)}^{2\tau}dw - \int_{\tau-2u}^{\tau}dw + \int_0^{\tau}dw  - 
\int_{\tau}^{2\tau}dw \right] G(n,u+w) \nonumber \\
&+& \frac{A}{\gamma^2}\int_{\tau}^{+\infty} du \, g(u)  \left[ \int_0^{\tau} dw -
\int_{\tau}^{2\tau}dw + \int_{0}^{2\tau}dw  - \int_{-\tau}^{\tau}dw \right] G(n,u+w) \nonumber \\
&=&\frac{A}{\gamma^2} \int_0^{\tau} du \, g(u)  \left[ \int_0^{\tau} dw +\int_0^{\tau-2u}
\!\!\!\!\!dw  \right] G(n,u+w) + \frac{A}{\gamma^2}\int_{\tau}^{+\infty} \!\!\!\! 
du \ g(u)  \left[ \int_0^{\tau} dw -\int_{-\tau}^{0}dw  \right] G(n,u+w) \nonumber \\
\end{eqnarray}   
\begin{eqnarray}
H_{ss}(n,\tau) &=& \frac{A}{\gamma^2}
\left\{ \int_0^{\tau} \!\!\!\!
du \, g(u)  \left[ F_n(\tau+u) - 2 F_n(u) + F_n(\tau-u)  \right]  + 
\int_{\tau}^{+\infty} 
\!\!\!\!\!\!\!\!\! du \, g(u)  \left[ F_n(\tau+u) - 2F_n(u) + F_n(u-\tau)  \right] 
\right\} \nonumber \\ 
\label{H_F}
\end{eqnarray}
which, after combining into a single term, corresponds to the result reported in Eq.~\eqref{Hss}, 
where we have defined $F_n (u) \equiv \int_0^u dw \, G(n,w)$. It is convenient to rewrite 
$F_n(\tau)$, using some variable transformations and integration by parts, as follows 
\begin{eqnarray}
    F_n(\tau) &=& \int_{0}^\tau \sqrt{\frac{\tau_0}{4\pi w}} \, e^{-\displaystyle{\frac{n^2 \tau_0}{4w}}} dw = 
    \frac{n \tau_0}{4\sqrt{\pi}} \int_{\alpha^2(\tau)}^{+\infty}  z^{-3/2} e^{-z} dz =
    \frac{n \tau_0}{2} \left[ \frac{e^{-\displaystyle{\alpha^2(\tau)}}}{ \sqrt{\pi} \alpha(\tau)}
    - \frac{2}{\sqrt{\pi}} \int_{\alpha(\tau)}^{+\infty} e^{-x^2} dx \right] 
    \nonumber \\
    &=&
    \frac{n \tau_0}{2} \left\{ \frac{e^{-\displaystyle{\alpha^2(\tau)}}}{ \sqrt{\pi} \alpha(\tau)}
    - \text{Erfc}[\alpha(\tau)] \right\}
\label{Fn_exp}
\end{eqnarray}
which is the result reported in Eq.~\eqref{def_Fn}.
In the transient protocol, we set ${\mathcal B}_n(\tau)=0$, hence,
\begin{eqnarray}
H_{tr}(n,\tau) &=& \langle {\cal A}_n(\tau) {\cal A}_0(\tau) \rangle 
\frac{A}{\gamma^2}
 \int_0^{\tau} \!\!\!\!
du \, g(u)  \left[ F_n(2\tau-u) -  F_n(u)  \right] 
\nonumber \\
\end{eqnarray}
which is the result reported in Eq.~\eqref{H_F_tr}.
\end{widetext}


%
\end{document}